\documentclass[%
 reprint,
 amsmath,amssymb,
 aps,
]{revtex4-2}

\usepackage{graphicx}
\usepackage{dcolumn}
\usepackage{bm}
\usepackage{here}
\usepackage{url}

\usepackage{amsmath}
\usepackage{multirow}
\usepackage{titlesec}

\allowdisplaybreaks[4]

\titlespacing*{\section}
  {0pt}{3ex plus 1ex minus .2ex}{2ex plus .2ex}
\titlespacing*{\subsection}
  {0pt}{3ex plus 1ex minus .2ex}{1ex plus .2ex}
\titlespacing*{\subsubsection}
  {0pt}{3ex plus 1ex minus .2ex}{1ex plus .2ex}

\begin{document}

\preprint{APS/123-QED}


\title{Data-Driven Discovery and Formulation Refines the Quasi-Steady Model of Flapping-Wing Aerodynamics}

\author{Yu Kamimizu}
  \affiliation{
    Graduate School of Science and Engineering, Chiba University, Chiba 263-8522, Japan
  }
\author{Hao Liu}
\author{Toshiyuki Nakata}
  \email{tnakata@chiba-u.jp}
  \affiliation{
    Graduate School of Engineering, Chiba University, Chiba 263-8522, Japan
  }
\date{\today}

\begin{abstract}
  Insects control unsteady aerodynamic forces on flapping wings to navigate complex environments. While understanding these forces is vital for biology, physics, and engineering, existing evaluation methods face trade-offs: high-fidelity simulations are computationally or experimentally expensive and lack explanatory power, whereas theoretical models based on quasi-steady assumptions offer insights but exhibit low accuracy. To overcome these limitations and thus enhance the accuracy of quasi-steady aerodynamic models, we applied a data-driven approach involving discovery and formulation of previously overlooked critical mechanisms. Through selection from 5,000 candidate kinematic functions, we identified mathematical expressions for three key additional mechanisms---the effect of advance ratio, effect of spanwise kinematic velocity, and rotational Wagner effect---which had been qualitatively recognized but were not formulated. Incorporating these mechanisms considerably reduced the prediction errors of the quasi-steady model using the computational fluid dynamics results as the ground truth, both in hawkmoth forward flight (at high Reynolds numbers) and fruit fly maneuvers (at low Reynolds numbers). The data-driven quasi-steady model enables rapid aerodynamic analysis, serving as a practical tool for understanding evolutionary adaptations in insect flight and developing bio-inspired flying robots.
\end{abstract}

\maketitle

\setlength{\extrarowheight}{4pt}
\setlength{\tabcolsep}{12pt}



%
\section{\label{sec:1_introduction}INTRODUCTION}

\begin{figure*}[t]
    \centering
    \includegraphics[width=1\textwidth]{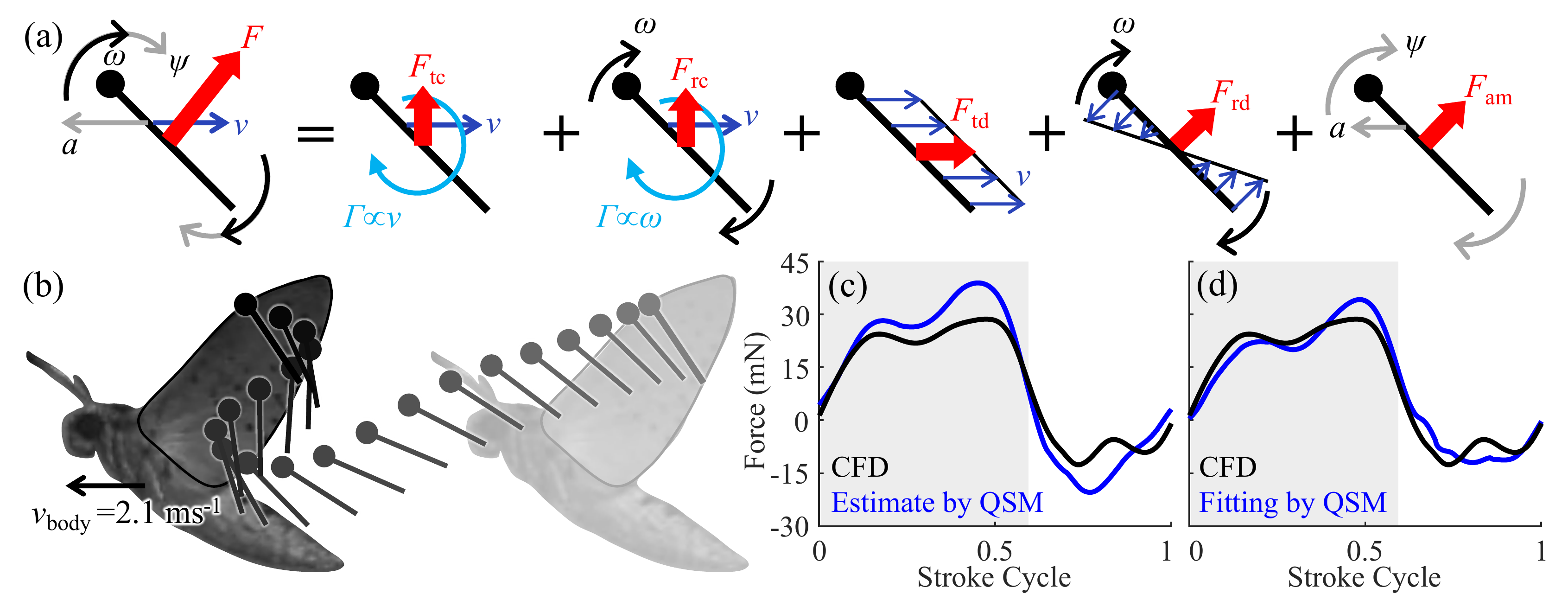}
    \caption{
      Conventional QSM and its accuracy. (a) Quasi-steady aerodynamic force (left) is composed of translational circulation (tc), rotational circulation (rc), translational drag (td), rotational drag (rd), and force due to added mass (am). (b) Side view of the wing kinematics of hawkmoth forward flight at 2.1 $\mathrm{ms^{-1}}$. The lines and circles represent the time series of the wing cross-sections (at 70\% of the wing length from the wing base) and leading edges, respectively. (c,d) Time series of the aerodynamic force normal to the wing surface. The blue lines represent the estimate by conventional QSM in (c) and the result of fitting of QSM in (d). The black lines represent the estimate by CFD as a ground truth. The gray shaded areas in (c,d) represent the downstroke.
    }
    \label{fig:10_conventional_QSM}
\end{figure*}

Insects exhibit exceptional flight capabilities by generating unsteady aerodynamic forces using their flapping wings. These abilities enable long-distance migrations, agile maneuvers in cluttered environments, and rapid escape responses beyond human reaction time \cite{Crall2015-nx,Johnson1969-ak,Muijres2014-hx}. Despite advances in modern technology, their efficient, stable, and maneuverable flight in unpredictable natural environments continues to outperform the state-of-the-art aerial robots.

The remarkable performance of insect flights has attracted the attention of researchers in biology, physics, and engineering. Understanding the aerodynamic mechanisms behind flapping flight not only addresses fundamental biological questions regarding wing and musculoskeletal adaptations \cite{Le_Roy2021-vr,Marden2000-gu,Aiello2021-xu,Gau2023-zm} but also inspires the development of innovative flying robots with novel unsteady aerodynamics \cite{Liu2016-um,Phan2020-el,Ma2013-qi,Floreano2009-tp}. However, the flapping flight of insects is fundamentally different from conventional propulsion using fixed or rotary wings, limiting the direct application of existing aerodynamic knowledge for understanding and replicating flight strategies.

Various approaches, including computational and robotic simulations and mathematical models, have been employed to investigate the mechanism by which insects generate aerodynamic forces for their survival \cite{Bomphrey2017-ys,Farisenkov2022-en,Nakata2023-pa,Poletti2024-fm,Dickinson1999-ug,Gravish2018-th,Ellington1984-ej,Xuan2020-ca}. Although computational fluid dynamics (CFD) provides detailed flow fields and force data with high accuracy, and robotic models offer insights into real-world physical phenomena, both approaches are computationally or experimentally expensive, restricting their use in studies requiring broad parameter exploration, such as systematic parameter sweeps or optimization. Moreover, these approaches primarily aim to reproduce phenomena with high fidelity; however, they do not explain the underlying mechanisms.

Mathematical models decompose complex aerodynamic phenomena into interpretable equations, enabling rapid performance evaluation across diverse conditions \cite{Ellington1984-ej,Xuan2020-ca,Nabawy2014-dr,Wang2016-ws,Ansari2006-xr}. Among these, the quasi-steady model (QSM) has been widely adopted owing to its balance between computational efficiency and ability to predict instantaneous aerodynamic forces \cite{Dickinson1999-ug,Xuan2020-ca,Wang2016-ws,Sane2001-gl,Wang2004-rp,Khan2005-zv,Hedrick2006-ji,Berman2007-tg,Dickson2008-ex,Bierling2009-nb,Truong2011-wu,Nakata2015-fr,Cheng2016-xj,Lee2016-hh,Armanini2016-qh,Cai2021-ve,Walker2021-zu,Osborne1951-dy,Ellington1984-li,Weis-Fogh1973-km,Sane2002-ch,Andersen2005-cl,van-Veen2023-dq,Sane2003-tm,Deng2006-rz,Whitney2010-fr}. The QSM assumes that aerodynamic forces depend solely on instantaneous wing attitude, velocity, and acceleration and expresses forces using a blade element approach, where the wing is divided into a series of blades, and the forces applied to each blade are integrated \cite{Osborne1951-dy,Ellington1984-li,Weis-Fogh1973-km}, considering five primary mechanisms as follows \cite{Wang2016-ws,Nakata2015-fr,Cai2021-ve} (Fig. \ref{fig:10_conventional_QSM}(a), Eq. \ref{eq:QSM_primary5}).
\begin{equation}\label{eq:QSM_primary5}
  F_\mathrm{aero} = F_{\mathrm{tc}} + F_{\mathrm{rc}} + F_{\mathrm{td}} + F_{\mathrm{rd}} + F_{\mathrm{am}}.
\end{equation}
$F_{\mathrm{tc}}$ and $F_{\mathrm{rc}}$ are the translational \cite{Dickinson1999-ug} and rotational \cite{Dickinson1999-ug,Sane2002-ch} circulatory lifts produced by circulation owing to translational or rotational motion (wing feathering motion), respectively. $F_{\mathrm{td}}$ and $F_{\mathrm{rd}}$ are the translational \cite{Dickinson1999-ug} and rotational \cite{Berman2007-tg,Andersen2005-cl} drags produced by translational and rotational motions, respectively. $F_{\mathrm{am}}$ is the force due to the added mass \cite{Osborne1951-dy,Ellington1984-yn}, that is, the force required to accelerate or decelerate the surrounding fluid. Each mechanism consists of a product of the morphological and kinematic terms. The incorporation of additional effects, such as the Wagner effect and the Reynolds number ($Re$) effect, improves the accuracy of the QSM in hovering flights \cite{Lee2016-hh,van-Veen2023-dq,Wagner1924-jb,Lentink2009-lx}. The coefficients of the circulatory force, or drag, are derived from simplified experiments, and more accurate QSMs can reportedly be obtained by adjusting the coefficients to fit the kinematics-force correlation obtained, for example, by high-fidelity CFD \cite{Nakata2015-fr,Cai2021-ve,Calado2023-la}.


Despite efforts to improve accuracy, errors are observed in conventional QSMs when applied to forward and maneuvering flights. When calculating aerodynamic forces on flapping wings during hawkmoth forward flight ($2.1 \mathrm{ms}^{-1}$ \cite{Willmott1997-ow}; Fig. \ref{fig:10_conventional_QSM}(b)), the conventional QSM with five mechanisms, Wagner effect, and $Re$ effect (based on instantaneous wing tip velocity; see Appendix \ref{sec:A2_App_convmodel} for derivation) show a 29\% error compared to the high-fidelity CFD results (Fig. \ref{fig:10_conventional_QSM}(c)). Even with QSM coefficients fitted to CFD data (see Appendix \ref{sec:A2_App_convmodel}), the waveform fails to reproduce the force fluctuations from the CFD results (Fig. \ref{fig:10_conventional_QSM}(d)). This critical limitation—which arises because QSMs are primarily developed for a hovering flight with wing rotation about the base—indicates that important aerodynamic mechanisms remain unaccounted for during forward and maneuvering flights. Studies on forward-flapping flights have focused on qualitative observations or simplified correlations \cite{Armanini2016-qh,Cai2021-ve,Walker2021-zu}, with few attempts to mathematically formulate the missing force-generation mechanisms. To the best of our knowledge, studies on the prediction of instantaneous aerodynamic forces during complex maneuvers have not yet been reported.

Recent advances in data-driven methodologies offer a promising approach to address this gap. Sparse identification of nonlinear dynamical systems (SINDy) \cite{Brunton2016-ld} has emerged as an efficient tool for discovering interpretable governing equations from data, outperforming traditional symbolic regression \cite{Bongard2007-cy,Schmidt2009-ju} in terms of computational cost. Therefore, SINDy has been used in a wide range of applications, such as fluid dynamics \cite{Fukami2021-vx,Foster2022-tp}, model predictive control \cite{Fasel2021-xt}, chemical reaction networks \cite{Hoffmann2019-lv} and predator-prey models \cite{Anjos2023-xq}. Given that the QSM expresses aerodynamic forces as linear combinations of nonlinear kinematic functions identically to the SINDy's formulation, we hypothesize that applying SINDy can reveal previously overlooked mechanisms in flapping-wing aerodynamics.

In this study, a data-driven model identification approach was applied using diverse wing kinematics and CFD-derived aerodynamic forces as training data. We identified the key functions representing the missing aerodynamic mechanisms that are critical for accurate force prediction, particularly in free flights with body motion. By integrating these mechanisms into the QSM, we demonstrated significant improvements in the prediction accuracy across various flight modes, including hovering, forward, and maneuvering flights, over a range of $Re$.

%
\section{\label{sec:2_methods}METHODS}
\begin{figure*}[tbp]
    \centering
    \includegraphics[width=1.0\textwidth]{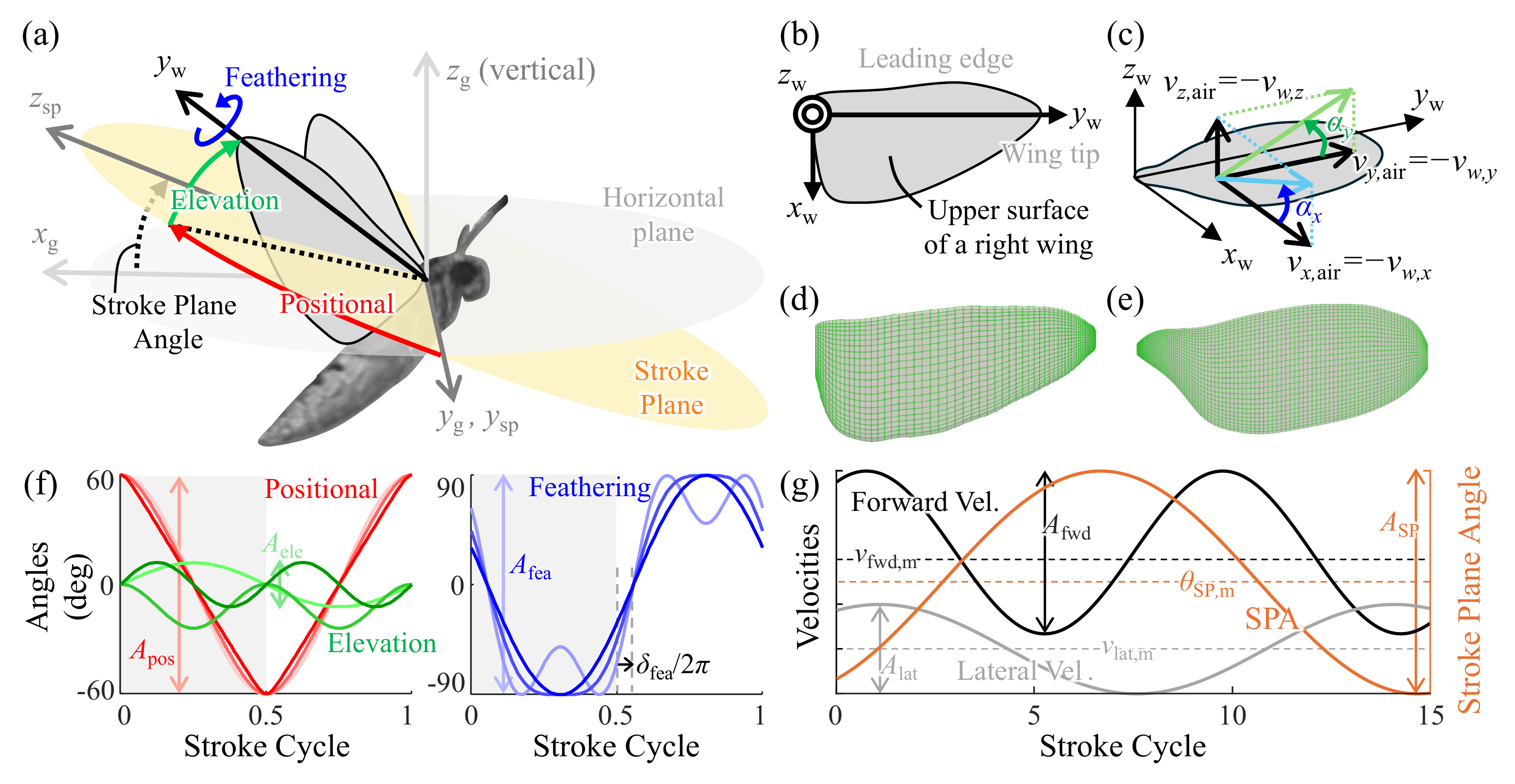}
    \caption{
      Wing kinematics and morphology. (a) Definition of the global coordinate system ($x_{\mathrm{g}}$, $y_{\mathrm{g}}$, $z_{\mathrm{g}}$), stroke-plane-fixed system ($x_{\mathrm{sp}}$, $y_{\mathrm{sp}}$, $z_{\mathrm{sp}}$), and flapping angles that define the wing orientation. (b) Wing-fixed coordinate system ($x_{\mathrm{w}}$, $y_{\mathrm{w}}$, $z_{\mathrm{w}}$). (c) Definition of the velocity ($v$) in the wing-fixed coordinate system and angle of attack ($\alpha$). (d,e) Wing surface mesh of the hawkmoth \cite{Aono2006-fu} (d) and fruit fly \cite{Muijres2014-hx} (e) for CFD analyses. (f) Variation of the flapping angles for the training data. $A_{\mathrm{pos}}$, $A_{\mathrm{ele}}$, $A_{\mathrm{fea}}$, and $\delta_{\mathrm{fea}}$ represent the amplitude of the positional, elevation, and feathering angles and the phase of the feathering angle with respect to the positional angle, respectively. The gray shaded areas represent the downstroke. (g) Variation of the forward flight velocity (black), lateral flight velocity (gray) and SPA (orange) for the training data. $A_{\mathrm{fwd}}$, $A_{\mathrm{lat}}$, $A_{\mathrm{SP}}$, $v_{\mathrm{fwd,m}}$, $v_{\mathrm{lat,m}}$, and $\theta_{\mathrm{SP,m}}$ represent the amplitude and average of the forward and lateral flight velocity and SPA, respectively.
    }
    \label{fig:20_wingkin_morphology}
\end{figure*}
%
\subsection{Wing kinematics \label{subsec:wingKinematics}}
In this study, three coordinate systems were defined, as shown in Figs. \ref{fig:20_wingkin_morphology}(a) and (b): global (inertial) frame ($x_{\mathrm{g}}$, $y_{\mathrm{g}}$, $z_{\mathrm{g}}$), stroke-plane-fixed frame ($x_{\mathrm{sp}}$, $y_{\mathrm{sp}}$, $z_{\mathrm{sp}}$), and wing-fixed frame ($x_{\mathrm{w}}$, $y_{\mathrm{w}}$, $z_{\mathrm{w}}$). The origin of the stroke plane-fixed frame was located at the center of mass of the body, whereas that of the wing-fixed frame was positioned at the wing pivot. Appendix \ref{sec:A1_App_nomenclature} summarizes the notation.

The flapping wing was assumed to be flat, rigid, and rotating around the wing pivot with respect to the stroke plane, whereas the body was translated and rotated in three dimensions. The stroke plane is defined by the regression of the wingtip path, and the stroke plane angle $\theta_{\mathrm{SP}}$ represents its horizontal inclination. The wing orientation relative to the stroke plane is described by three flapping angles: positional, elevation, and feathering angles (Fig. \ref{fig:20_wingkin_morphology}(a)). The positional angle $\theta_{\mathrm{pos}}$ and elevation angle $\theta_{\mathrm{ele}}$ represent the azimuth and elevation of the wing tip with respect to the stroke plane, respectively. The feathering angle $\theta_{\mathrm{fea}}$ describes pronation or supination around the wing's longitudinal (spanwise) axis $y_{\mathrm{w}}$. The coordinate transformation matrix ${\bf R}_{\mathrm{g,w}}$ which converts a vector from the global frame to a wing-fixed frame, is expressed as:
\begin{equation}
  {\bf R}_{\mathrm{g,w}} = {\bf R}_2(\pi/2 - \theta_{\mathrm{SP}}) {\bf R}_1(\theta_{\mathrm{pos}}) {\bf R}_3(\theta_{\mathrm{ele}}) {\bf R}_2(\theta_{\mathrm{fea}}),
\end{equation}
where ${\bf R}_1(\theta)$, ${\bf R}_2(\theta)$, and ${\bf R}_3(\theta)$ are the rotation matrices around the $x$-, $y$-, and $z$-axes, respectively. The angular velocity of the wing relative to the body $\bm{\omega}_{\mathrm{b,w}}$ (expressed in the wing-fixed frame) is computed by
\begin{align}
  \bm{\omega}_{\mathrm{b,w}}
  &=
  \begin{bmatrix}
      0\\
      \dot{\theta}_{\mathrm{fea}} \\ 
      0 \\
  \end{bmatrix}
  +{\bf R}_2(\theta_{\mathrm{fea}})^\mathrm{T}
  \begin{bmatrix}
      0\\
      0 \\
      \dot{\theta}_{\mathrm{ele}} \\ 
  \end{bmatrix}\nonumber \\ 
  &\quad +
  {\bf R}_2(\theta_{\mathrm{fea}})^\mathrm{T}
  {\bf R}_3(\theta_{\mathrm{ele}})^\mathrm{T}
  \begin{bmatrix}
      \dot{\theta}_{\mathrm{pos}} \\ 
      0\\
      0 \\
  \end{bmatrix} \nonumber \\ 
  &\quad +
  {\bf R}_2(\theta_{\mathrm{fea}})^\mathrm{T}
  {\bf R}_3(\theta_{\mathrm{ele}})^\mathrm{T}
  {\bf R}_1(\theta_{\mathrm{pos}})^\mathrm{T}
  \begin{bmatrix}
      0\\
      \dot{\theta}_{\mathrm{SP}} \\ 
      0 \\
  \end{bmatrix}.
\end{align}
Similarly, the angular velocity of the body relative to the global frame $\bm{\omega}_{\mathrm{b}}$ is computed from the roll, pitch, and yaw angle of the body. The angular velocity of the wing relative to the global frame is given by 
\begin{equation}
  \bm{\omega}_{\mathrm{w}} = \bm{\omega}_{\mathrm{b}} + \bm{\omega}_{\mathrm{b,w}}.
\end{equation}
The angular acceleration $\psi_{\mathrm{w}}$ is expressed as
\begin{equation}
  \bm{\psi}_{\mathrm{w}}=
  {\bf R}_{\mathrm{g,w}}^\mathrm{T} 
  \frac{\mathrm{d} ({\bf R}_{\mathrm{g,w}} \bm{\omega}_{\mathrm{w}})}{\mathrm{d} t} .
\end{equation}
The velocity ${\bf v}_{\mathrm{w}}$ of a point ${\bf p}$ on the wing in the wing-fixed frame is given by:
\begin{equation}
  {\bf v}_{\mathrm{w}} =
  {\bf v}_{\mathrm{b}} + 
  \bm{\omega}_{\mathrm{w}} \times {\bf p},
\end{equation}
where ${\bf v}_{\mathrm{b}}$ is the translational velocity of the body (or origin of the stroke plane) in the wing-fixed frame. Thus, the acceleration of the wing ${\bf a}_{\mathrm{w}}$ can be written as
\begin{equation}
  {\bf a}_{\mathrm{w}}=
  {\bf R}_{\mathrm{g,w}}^\mathrm{T}
  \frac{\mathrm{d} ({\bf R}_{\mathrm{g,w}} \bm{v}_{\mathrm{w}})}{\mathrm{d} t}.
\end{equation}
The angles of attack in the chordwise and spanwise directions (Fig. \ref{fig:20_wingkin_morphology}(c)) can be calculated as
\begin{align}
  \alpha_x &= \arctan
  \left( \frac{-v_{\mathrm{w},z}}{-v_{\mathrm{w},x}} \right) \notag \\
  \alpha_y &= \arctan
  \left( \frac{-v_{\mathrm{w},z}}{-v_{\mathrm{w},y}} \right) 
\end{align}
respectively, where 
$v_{\mathrm{w},x}$, $v_{\mathrm{w},y}$, and $v_{\mathrm{w},z}$
are the $x_{\mathrm{w}}$-, $y_{\mathrm{w}}$-, and $z_{\mathrm{w}}$-components of ${\bf v}_{\mathrm{w}}$. Appendix \ref{sec:A1_App_nomenclature} summarizes the notation.

%
\subsection{Data-driven model discovery \label{subsec:modelDiscovery}}
We utilized a data-driven model discovery method to identify a set of kinematic functions that effectively explains the variations in the aerodynamic forces. First, a wing kinematics dataset covering conditions from hovering to free flight was constructed for two differently sized insects to cover a range of $Re$: the hawkmoth (wing length of $\sim 50$ mm; $Re \sim 10^4$) and fruit fly (wing length of $\sim 3$ mm; $Re \sim 10^2$). The corresponding aerodynamic forces were computed using a CFD analysis. The correlation between the kinematic and aerodynamic forces served as the training data for the data-driven model discovery method. Second, as an initial step in model construction, a conventional QSM was formulated based on previous studies \cite{Dickinson1999-ug,Wang2016-ws,Berman2007-tg,Nakata2015-fr,Cai2021-ve}. Subsequently, from a library of kinematic functions, we identified potentially important functions for aerodynamic force estimation using SINDy, which is a sparse identification method for model discovery. Finally, we attempted to interpret and discover additional aerodynamic mechanisms that contribute to force production and are overlooked by conventional QSM.

\setlength{\tabcolsep}{4pt}
\begin{table}[tbp]
  \caption{\label{tab:1_param_cfd} Parameters for computational fluid dynamic analyses.}
  \begin{ruledtabular}
    \begin{tabular}{lcc}
      & Hawkmoth & Fruit fly \\
      \hline
      Wing length (mm) & 51.5 & 3.20 \\
      Mean chord length (mm) & 18.3 & 0.97 \\
      Flapping frequency (Hz) & 25 & 188 \\
      Reynolds number & 4300--16000 & 110--280 \\
      Wing grid & $45 \times 65 \times 101$ & $55 \times 110 \times 101$ \\
      Dimensionless timestep & 0.01 & 0.01 \\
    \end{tabular}
  \end{ruledtabular}
\end{table}
\setlength{\tabcolsep}{12pt}

%
\subsubsection{Training data \label{subsubsec:trainingData}}
The wing kinematics datasets for hawkmoths and fruit flies were generated by systematically sweeping the waveforms of the translations and rotations of both the wings and bodies, thereby covering a wide range of kinematic conditions (Figs. \ref{fig:20_wingkin_morphology}(f) and (g)). The amplitudes, shapes, and phase differences of the waveforms for the flapping angles (i.e., positional, elevation, and feathering angles) were varied systematically (Fig. \ref{fig:20_wingkin_morphology}(f)). Among these, we optimized the selection of 50 wingbeats for the training dataset to ensure that the distributions of angular velocities and accelerations were as uniform as possible (see Appendix \ref{subsec:App_wingKinDataset} for details). To account for the body translation in three directions, forward velocity $v_{\mathrm{fwd}}$, lateral velocity $v_{\mathrm{lat}}$, and stroke plane angle $\theta_{\mathrm{SP}}$ were also varied in longer timescales than the period of wingbeats (Fig. \ref{fig:20_wingkin_morphology}(g)). These oscillations are centered at the mean forward velocity $v_{\mathrm{fwd,m}}$, mean lateral velocity $v_{\mathrm{lat,m}}$, and mean stroke plane angle $\theta_{\mathrm{SP,m}}$, with amplitudes $A_{\mathrm{fwd}}$, $A_{\mathrm{lat}}$, and $A_{\mathrm{SP}}$, and phase shift $\delta_{\mathrm{b}}$. Based on the kinematic correlation observed from insect flight, which suggests that the insects generate thrust by redirecting the force normal to the stroke plane, $\theta_{\mathrm{SP,m}}$ was formulated as a function of the mean forward velocity $v_{\mathrm{fwd,m}}$ (see Appendix \ref{subsec:App_wingKinDataset} for detail). $v_{\mathrm{fwd,m}}$ was determined so as to cover the flight speed of each insect. $v_{\mathrm{lat,m}}$, $A_{\mathrm{fwd}}$, and $A_{\mathrm{SP}}$ were held constant, and $A_{\mathrm{lat}}$ was set to 0.5$v_{\mathrm{fwd,m}}$. The angular velocity of the body $\bm{\omega}_{\mathrm{b}}$ was set to zero under the assumption that the variations in $\theta_{\mathrm{pos}}$, $\theta_{\mathrm{ele}}$, and $\theta_{\mathrm{fea}}$ sufficiently represent the influence of the body rotation. By varying $v_{\mathrm{fwd,m}}$ over four values and $\delta_{\mathrm{b}}$ over six values, and applying 50 wingbeats for each body motion, the complete kinematics dataset comprised 1,200 wingbeats.

Numerical simulations were performed using a versatile high-fidelity CFD simulator to compute the aerodynamic forces corresponding to a wing kinematic dataset \cite{Liu2009-xk}. This simulator integrates the modeling of realistic wing morphology, wing kinematics, and the unsteady aerodynamics of insect flight. In the simulator, the three-dimensional, incompressible, and unsteady Navier-Stokes equations (written in a strong conservation form for mass and momentum) were solved using the finite volume method. The equations were converted into a semi-discrete form by introducing the generalized Reynolds transport theorem and employing the Gauss integration theorem and were discretized in time using a Padé scheme. Because this solver has been well-validated \cite{Liu2009-xk,Nakata2012-og} through comparisons with other experimental \cite{Dickinson1999-ug,Heathcote2007-og,Heathcote2008-xn,Lua2010-vj} or numerical \cite{Sun2002-eg,Chimakurthi2009-ej,Dai2012-vp,Le2013-zl,Wan2015-fh} estimates of aerodynamic forces across a wide range of $Re$ values ($10^1$--$10^4$), we adopt data from the simulator as the ground truth. Further details of the solver are available in a previous report \cite{Liu2009-xk}. For our simulation, we used the wing morphologies of the hawkmoth \cite{Aono2006-fu} and fruit fly \cite{Muijres2014-hx} (Figs. \ref{fig:20_wingkin_morphology}(d) and (e); see Appendix \ref{subsec:App_validationCFD} for the validation and verification) and simulated the flow fields around a single wing, excluding the body, for simplicity. The parameters of the wing shape, $Re$, and computational conditions are summarized in Table \ref{tab:1_param_cfd}. $Re$ is defined as $U_{\mathrm{tip}} c_{\mathrm{m}} / \nu$, where $U_{\mathrm{tip}}$ ($2 \Phi R f + v_{\mathrm{body}}$, where $\Phi, R, f$, and $v_{\mathrm{body}}$ are the wingbeat amplitude, wing length, wingbeat frequency, and body velocity, respectively) is the mean wingtip velocity, $c_{\mathrm{m}}$ is the mean chord length, and $\nu$ is the kinetic viscosity of the surrounding fluid.

%
\subsubsection{Conventional QSM of flapping-wing aerodynamics \label{subsubsec:conventionalQSM}}
As an initial model, a conventional QSM was formulated based on previous studies that adopted for hovering, forward flight, and free flight \cite{Dickinson1999-ug,Wang2016-ws,Berman2007-tg,Nakata2015-fr,Cai2021-ve}. In this model, the aerodynamic force was computed using the blade element method (BEM), in which the wing was divided into chordwise blades with an infinitesimal span length $\mathrm{d}y$ (referred to as blade elements). The aerodynamic forces acting on the wings were calculated by integrating those of all the the elements. The conventional QSM accounts for the contributions of the five primary mechanisms (as introduced in Sec. \ref{sec:1_introduction}) to the force acting on a blade element located at a spanwise position $y$ from the wing pivot.
\begin{align}
  \mathrm{d} F_{\mathrm{tc}} 
  &=
  \frac{1}{2} \rho v_{\mathrm{w},xz}^2
  C_{L}(\alpha_x) \cos(\alpha_x) c \mathrm{d}y, \label{eq:convQSM_simple_tc} \\
  \mathrm{d} F_{\mathrm{td}} 
  &=
  \frac{1}{2} \rho v_{\mathrm{w},xz}^2
  C_{D}(\alpha_x) \sin(\alpha_x) c \mathrm{d}y, \label{eq:convQSM_simple_td} \\
  \mathrm{d} F_{\mathrm{rc}}
  &=
  - \rho v_{\mathrm{w},x} \beta_{\mathrm{rc}} \omega_{\mathrm{w},y} c^2 \mathrm{d}y, \label{eq:convQSM_simple_rc} \\
  \mathrm{d} F_{\mathrm{rd}}
  &=
  \frac{1}{2} \rho \omega_{\mathrm{w},y} |\omega_{\mathrm{w},y}| 
  \int _{x_{\mathrm{TE}}} ^{x_{\mathrm{LE}}} x|x| \mathrm{d}x ~ \beta_{\mathrm{rd}} \mathrm{d}y, \label{eq:convQSM_simple_rd} \\
  \mathrm{d} F_{\mathrm{AM}}
  &=
  - \beta_{\mathrm{am,trans}} \frac{\pi}{4} \rho c^2 a_{\mathrm{w},z} \mathrm{d}y, \notag \\
  &\quad +
  \beta_{\mathrm{am,rot}} \frac{\pi}{4} \rho c^2 \psi_{\mathrm{w},y} \frac{ x_{\mathrm{TE}} + x_{\mathrm{LE}}}{2} \mathrm{d}y, \label{eq:convQSM_simple_am}
\end{align}
where $\beta_{\mathrm{rc}}$, $\beta_{\mathrm{rd}}$, $\beta_{\mathrm{am,trans}}$, and $\beta_{\mathrm{am,rot}}$ are constant values depending on the wing morphology, $\rho$ is density of air, $c$ is chord length, 
$v_{\mathrm{w},xz} = \sqrt{v_{\mathrm{w},x}^2 + v_{\mathrm{w},z}^2}$,
$\omega_{\mathrm{w},y}$ is the $y_{\mathrm{w}}$-component of the wing's angular velocity, and $x_\mathrm{LE}$ and $x_\mathrm{TE}$ are the chordwise locations of the leading and trailing edges in the wing-fixed frame, respectively. Although the translational lift coefficient $C_L(\alpha_x)$ is often formulated as $\sin(2\alpha_x)$ following \cite{Dickinson1999-ug}, we adopted the formulation in \cite{Nakata2015-fr} to account for the peak shift of the lift coefficient by the downwash:
\begin{equation}
  C_{L}(\alpha) 
  =
  \beta_{\mathrm{tc,1}} \alpha_x \left(
    \alpha_x - \frac{\pi}{2}
  \right) +
  \beta_{\mathrm{tc,2}} \alpha_x^2 \left(
    \alpha_x - \frac{\pi}{2}
  \right),
\end{equation}
where $\beta_{\mathrm{tc,1}}$ and $\beta_{\mathrm{tc,2}}$ are coefficients determined by fitting. The translational drag coefficient $C_D(\alpha_x)$ is defined as $\beta_{\mathrm{td}} \sin^2(\alpha_x)$. Additionally, the model incorporated the Wagner effect and $Re$ effect based on previous studies \cite{Lee2016-hh,van-Veen2023-dq,Lentink2009-lx} (Appendix \ref{sec:A2_App_convmodel}, \ref{subsec:App_formulationWagner}). Note that, as the analyses were performed with a single rigid wing, we did not consider additional intricate mechanisms, such as the effects of the body, wing flexibility, and wing-wing interactions.

Because the velocity vector and angle of attack vary across the wing surface, calculating the aerodynamic forces acting on the entire wing during free flight requires numerical integration of these variables \cite{Lee2016-hh,Armanini2016-qh,Cai2021-ve,Walker2021-zu}, which is computationally expensive. In this study, the force acting on the entire wing surface was estimated based on the velocity and angle of attack of the representative points, which were optimized for each force production mechanism (see Appendix \ref{sec:A2_App_convmodel}), thereby enhancing the computational efficiency.

%
\begin{figure}[tbp]
    \centering
    \includegraphics[width=0.48\textwidth]{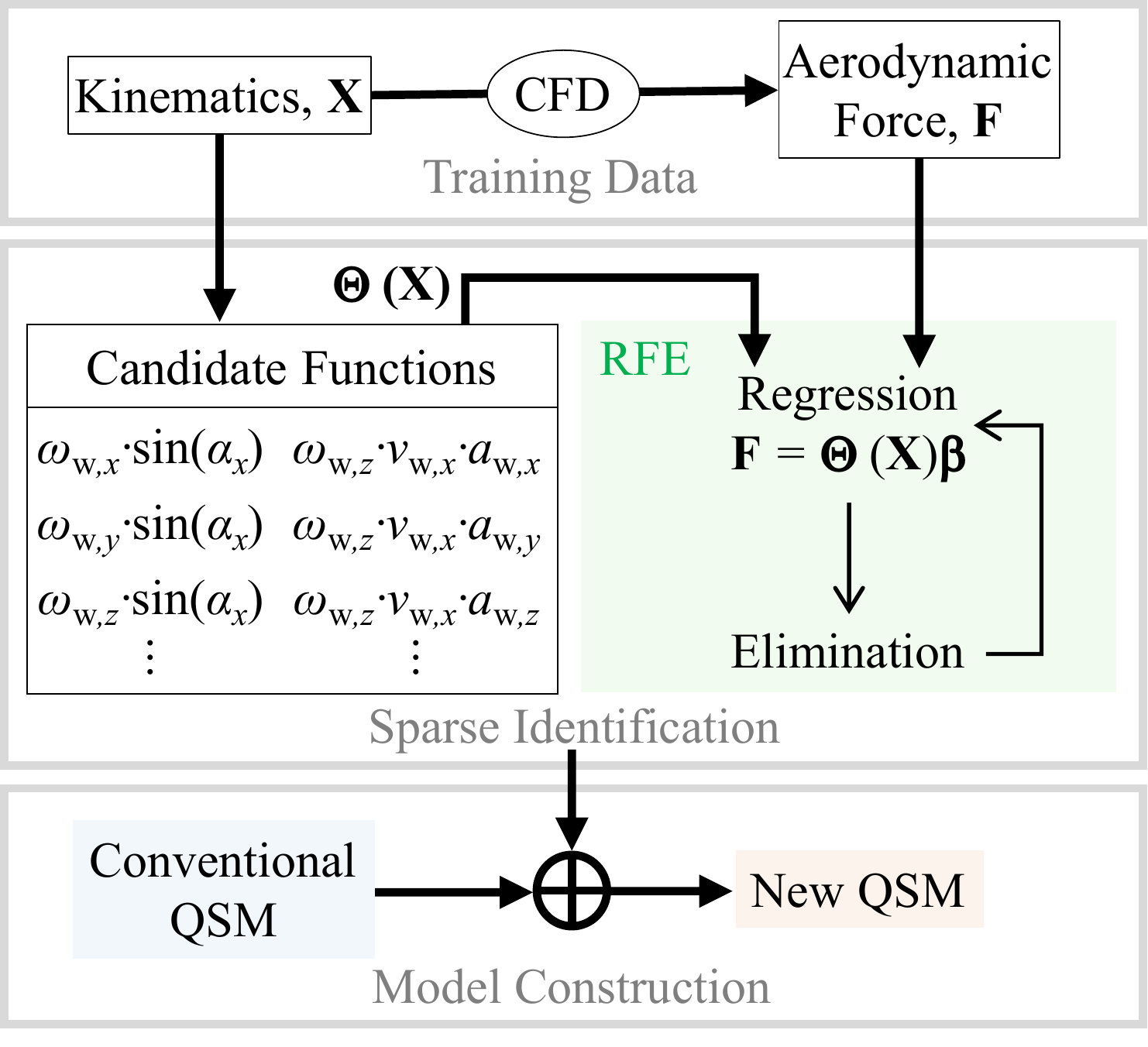}
    \caption{
      Construction of the data-driven QSM. The correlations between the wing kinematics and aerodynamic forces, used as training data, were obtained via CFD analyses (top). A list of candidate functions (middle left) was prepared by the combination of kinematic terms, from which important functions were selected using RFE (middle right). After careful examination, the identified mechanisms were integrated into the conventional QSM to construct the new QSM (bottom).
    }
    \label{fig:30_construction_DDQSM}
\end{figure}

\begin{figure*}[tbp]
    \centering
    \includegraphics[width=1.0\textwidth]{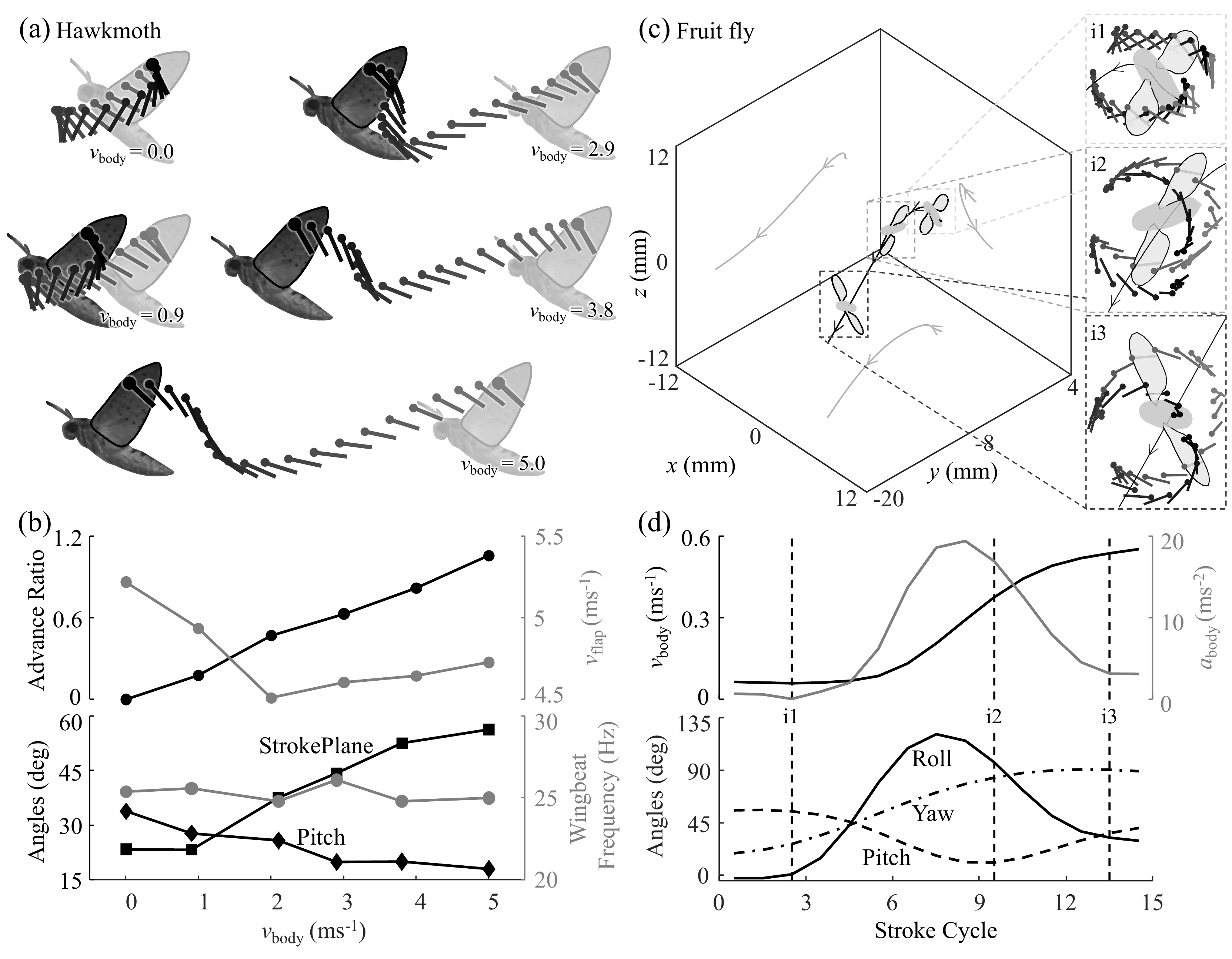}
    \caption{
      Kinematics data of the hawkmoth (a, b) and fruit fly (c, d) used for validation of the data-driven QSM. (a) Wing kinematics of the hawkmoth during hovering and forward flight \cite{Willmott1997-ow}. The case of forward flight at $2.1 \mathrm{ms^{-1}}$ is shown in Fig. \ref{fig:10_conventional_QSM}(b). (b) (top) advance ratio (black) and mean wingtip speed due to flapping (gray) and (bottom) body pitch angle (black diamonds), SPA (black squares), and wingbeat frequency (gray circles) of the hawkmoth at different flight speeds. (c) Body trajectory and wing kinematics of the fruit fly during maneuvering flight. The wing kinematics at i1 (slow forward flight), i2 (turning), and i3 (fast forward flight) in (d) are magnified on the right. (d) Time series of (top) body velocity (black) and acceleration (gray) and (bottom) roll (solid), pitch (dashed), and yaw (dash-dot) angles of the body of the fruit fly. The lines and circles in (a,b) represent the wing cross-sections (at 70\% of the wing length from the wing base) and leading edges, respectively.
    }
    \label{fig:40_kinematics_data_HMFF}
\end{figure*}

\subsubsection{Data-driven model discovery \label{subsubsec:modelDiscovery}}

We explored the kinematic functions corresponding to the additional aerodynamic mechanisms overlooked by conventional QSM using the training data described in Sec. \ref{subsubsec:trainingData} (top of Fig. \ref{fig:30_construction_DDQSM}). We constructed a library of candidate functions (middle of Fig. \ref{fig:30_construction_DDQSM}) comprising first-, second-, and third-order polynomials of the kinematic variables (velocity, acceleration, angular velocity, angular acceleration, and angle of attack; see Appendix \ref{subsec:App_candidateFuncs} for details). From this library, we identified potentially important functions for aerodynamic force estimation using SINDy \cite{Brunton2016-ld}. SINDy can simplify the model discovery problem to a linear regression problem by assuming that the governing equation or target function (the aerodynamic force in this study) can be expressed as a sparse linear combination of basis functions (kinematic functions in this study) from a predefined function library. This sparsity-promoting regression yields an interpretable mathematical expression that accurately describes the target function \cite{Brunton2016-ld,Rudy2017-fh,Champion2019-wy}.

\begin{figure}[tbp]
    \centering
    \includegraphics[width=0.48\textwidth]{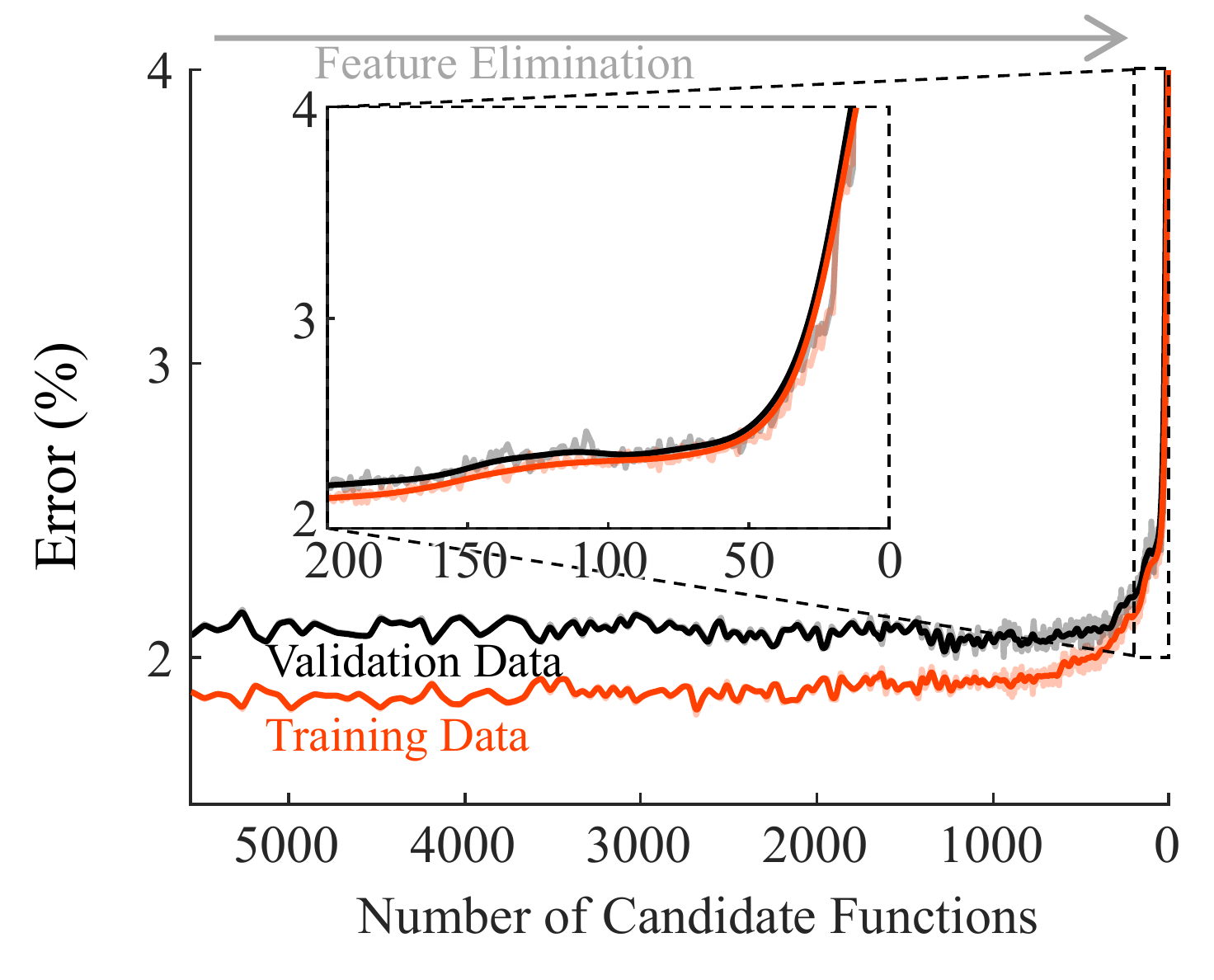}
    \caption{
      Effect of the number of candidate functions on the model performance. The plot shows the fitting error for the training data (orange) and the prediction error for the validation data (black). The thin lines represent the raw data, and the thick lines represent the smoothed results. A magnified view of the curves with fewer than 200 candidate functions is shown in the inset.
    }
    \label{fig:50_candidate_num_vs_performance}
\end{figure}

When too many candidate functions exist, the regression problem becomes numerically challenging, and the amount of data that can be used in a single regression is limited. To overcome these issues, STLS \cite{Brunton2016-ld} and STRidge \cite{Rudy2017-fh} adopt an approach to eliminate the less important candidate functions recursively based on the regression results until both the desired sparsity and precision are satisfied. These methods are similar to the recursive feature elimination (RFE) \cite{Guyon2002-ws} developed for gene selection and commonly used for statistical modeling, especially in areas where datasets contain many features. In this study, we adopted a recursive approach similar to STRidge \cite{Rudy2017-fh} and repeated the following operations (middle of Fig. \ref{fig:30_construction_DDQSM}):
\begin{enumerate}
  \item Perform RIDGE regression, a regularized linear regression method \cite{Hoerl1970-kp}, with the aerodynamic force as the target variable and the candidate functions as explanatory variables.
  \item Compute an importance index (see Appendix \ref{subsec:App_featureelim}) for each candidate function based on the resultant regression coefficients and eliminate functions with lower importance.
\end{enumerate}
After iterating these steps, the remaining candidate functions were considered potentially significant. We then attempted to interpret their aerodynamic implications, identify the underlying mechanisms, and construct a new data-driven QSM by incorporating these mechanisms (bottom of Fig. \ref{fig:30_construction_DDQSM}; see Appendix \ref{subsec:App_featureelim} for more details). Note that, although we present the results of our attempt to discover mechanisms using the hawkmoth dataset, we confirmed the similar results with fruit fly dataset.

The accuracy of the data-driven QSM was evaluated by comparing the estimates with the CFD results for hawkmoths ($Re \sim 10^4$) and fruit flies ($Re \sim 10^2$). The kinematic data of hawkmoths were adopted from \cite{Willmott1997-ow} (Moth F2), which includes hovering and forward flight data at various flight speeds (Figs. \ref{fig:40_kinematics_data_HMFF}(a) and (b)). For fruit flies, wing kinematics during evading maneuvers \cite{Muijres2014-hx} were used, which included a sequence of slow forward flight, turning, and subsequent forward flight (Figs. \ref{fig:40_kinematics_data_HMFF}(c) and (d)). The model coefficients were determined using the Levenberg-Marquardt algorithm through the minimization of a regularized objective function consisting of the squared error and an L2 penalty term to prevent overfitting, analogous to ridge regression (Appendix \ref{sec:A2_App_convmodel}), with the same data as employed for sparse identification. The aerodynamic forces were estimated based on the wing kinematics of each insect. The estimation errors were quantified as the root-mean-square error (RMSE) normalized by half the amplitude of the force variation estimated using CFD. Because the force parallel to the wing surface is an order of magnitude smaller than the force normal to the wing ($F_z$), only the evaluation results for the dominant force $F_z$ are presented unless otherwise noted. Note that the measured kinematic data used for model evaluation were not included in the training data for either sparse identification or tuning of coefficients.

\begin{table}[tbp]
	\caption{\label{tab:4_predrank_partial}%
	List of candidate functions suggested by sparse identification and extracted manually as physically interpretable functions. The numbers in the first column represent the ranking. The table of all the 100 candidate functions suggested by sparse identification are shown in supplementary material \cite{supplement}.
	}
	\begin{ruledtabular}
		\begin{tabular}{lc}
			3* & $ \sin(\alpha_y)  \cdot  v_{\mathrm{b},z}  \cdot  v_{\mathrm{b},z} $ \\
			10* & $ \omega_{\mathrm{w},x}  \cdot  v_{\mathrm{w},y} $ \\
			11* & $ \sin(\alpha_y)  \cdot  \omega_{\mathrm{w},y}  \cdot  v_{\mathrm{w},z} $ \\
			35* & $ \sin(\alpha_x)  \cdot  \omega_{\mathrm{w},z}  \cdot  v_{\mathrm{b},x} $ \\
			42* & $ \omega_{\mathrm{w},y}  \cdot  \mathrm{sign}(a_{\mathrm{w},x}) \sqrt{|a_{\mathrm{w},x}|} $ \\
			43* & $ \omega_{\mathrm{w},x}  \cdot  v_{\mathrm{b},y} $ \\
			64* & $ \sin(\alpha_x)  \cdot  \omega_{\mathrm{w},x}  \cdot  v_{\mathrm{w},z} $ \\
			74* & $ \sin^3(\alpha_y)  \cdot  \omega_{\mathrm{w},y}  \cdot  v_{\mathrm{b},z} $ \\
			78* & $ \sin(\alpha_x)  \cdot  v_{\mathrm{b},z}  \cdot  v_{\mathrm{w},z} $ \\
		\end{tabular}
	\end{ruledtabular}
\end{table}

%
\section{\label{sec:3_results}RESULTS}
The data-driven algorithms identified several important candidate functions. We attempted to interpret these functions and identified three aerodynamic mechanisms that have not been considered in conventional QSMs but contribute to aerodynamic force estimation. A new data-driven QSM was constructed considering these mechanisms. By comparing the accuracy of the conventional model (the model used in Figs. \ref{fig:10_conventional_QSM}(c) and (d) or Eq. \ref{eq:conventional_QSM} in Appendix \ref{sec:A2_App_convmodel}) and the new QSMs with respect to the estimation of aerodynamic forces using the CFD results as the ground truth, the importance of the three identified mechanisms and the high accuracy of the new QSM are demonstrated.

%
\begin{figure*}[t]
    \centering
    \includegraphics[width=1.0\textwidth]{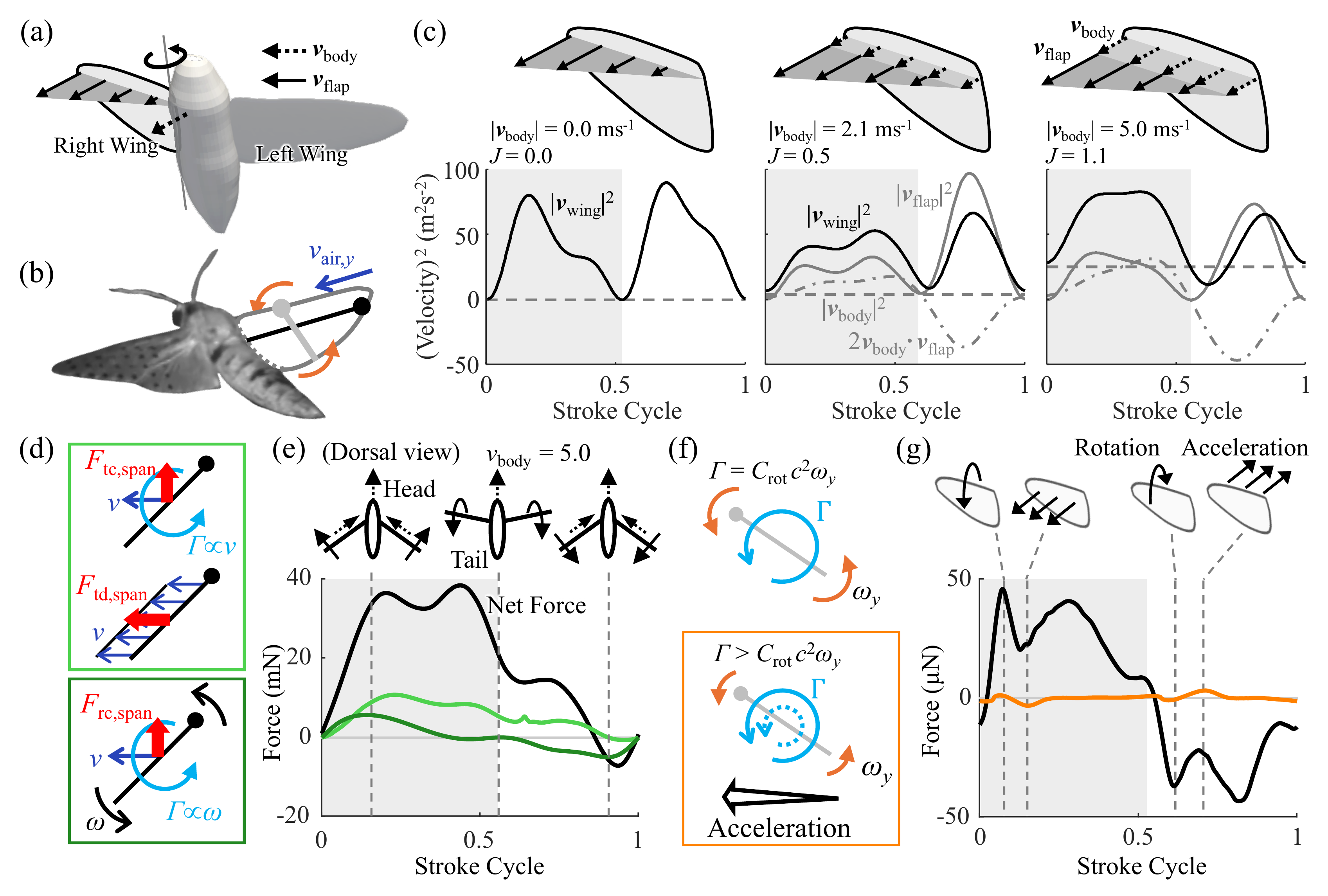}
    \caption{
      Three newly identified mechanisms discovered by sparse identification. (a) Velocities generated by body motion and flapping motion. (b) Air velocity due to spanwise motion (blue arrow) and the definition of the spanwise blade element (black line and circle). (c) Effects of the advance ratio on the distribution of the surface velocity (top) and (squared) wing velocity throughout a wingbeat (bottom). From left to right, the panels show the hawkmoth forward flight at $v_{\mathrm{body}} = 0$ (hovering; advance ratio $J = 0$), 2.1 ($J = 0.5$) and 5.0 $\mathrm{ms^{-1}}$ ($J = 1.1$). (d) Schematic of the aerodynamic force (red) induced by the air velocity relative to the spanwise blade elements (blue). (e) (top) Top view of the body and spanwise blade during the hawkmoth forward flight at $v_{\mathrm{body}} = 5.0 \mathrm{ms^{-1}}$. Time series of the spanwise translational force (light green), spanwise rotational force (dark green), and total aerodynamic force (black) (bottom). (f) Schematic of the generation of circulation due to wing rotation (top). 
      When the wing rotation slows down (bottom), the wing acceleration delays its decay, resulting in a larger circulation (solid circle) than the quasi-steady estimate (dotted circle).
      (g) Wing kinematics during the fruit fly forward flight at $v_{\mathrm{body}} = 0.54 \mathrm{ms^{-1}}$ (top; i3 in Figs. 4(c) and (d)) and the time series of the rotational Wagner effect (orange) and total aerodynamic force (black) (bottom). The gray shaded areas in (c,e,g) represent the downstroke.
    }
    \label{fig:60_new_mechanisms_explanation}
\end{figure*}

%
\subsection{Sparse identification \label{subsec:sparseIdentification}}
The addition of candidate functions reduced the error of the QSM compared with those of the conventional QSM, and the error increased as the number of candidate functions decreased (Fig. \ref{fig:50_candidate_num_vs_performance}). In particular, when the number of candidate functions fell below 50, the error increased rapidly, approaching the magnitude of errors of the conventional QSM ($\sim 5$\%). While the number of functions can be increased, and the functions can be used as a black box, we consider it important that each function is physically accountable to avoid overfitting, as most of the selected functions may only explain \textit{ad hoc} correlations specific to the system used for training. Therefore, we manually examined 100 selected functions that were considered large enough to include important functions (Fig. \ref{fig:50_candidate_num_vs_performance}) and were barely small enough to allow manual verification of their physical interpretation. We repeated the algorithm 10 times, and the same 100 functions were selected every time (see Table \ref{tab:4_predrank_partial}), despite the random process involved. Therefore, we considered these 100 selected functions as robust and important candidates for new mechanisms, carefully examined them, and decided to use only those terms that could be explained physically in the new QSM, as described below.

\subsection{Identified aerodynamic mechanisms \label{subsec:identifiedMechanisms}}
After a close examination of the 100 selected candidate functions, three mechanisms were identified: the effect of the advance ratio, effect of spanwise kinematic velocity, and rotational Wagner effect (Fig. \ref{fig:60_new_mechanisms_explanation}). Each of these mechanisms has been identified as important in previous studies but has not been formulated. We explain the details of these mechanisms through mathematical expressions in this section. The accuracy of QSM is discussed in Sec. \ref{subsec:improvementInAccuracy}.

The advance ratio, that is, the ratio of the velocity of the body to that of the wingbeat, affects the spatial velocity distribution on the wing surface and must be considered when calculating the aerodynamic forces generated by the flapping wing translating with the body (Fig. \ref{fig:60_new_mechanisms_explanation}(a)). In conventional QSMs, translational lift and drag are expressed by an equation that combines the velocity of the body (i.e., the translational and rotational velocities of the body) and the velocity of the wingbeat (i.e., the velocity due to the flapping of the wings) as follows \cite{Dickson2008-ex,Truong2011-wu,Armanini2016-qh,Cai2021-ve,Deng2006-rz}:
\begin{align}
  \mathrm{d} F
  &\propto
  \| {\bf v}_{\mathrm{wing}} \|^2
  C_{F}(\alpha) \notag \\
  &=
  \| {\bf v}_{\mathrm{body}} + {\bf v}_{\mathrm{flap}} \|^2
  C_{F}(\alpha), \label{eq:convQSM_tctd}
\end{align}
which can be decomposed as
\begin{align}
  \mathrm{d} F 
  &\propto
  \| {\bf v}_{\mathrm{body}} \|^2
  C_{F,\mathrm{bd}}(\alpha) \notag \\
  &\quad +
  \| {\bf v}_{\mathrm{flap}} \|^2
  C_{F,\mathrm{fl}}(\alpha) \notag \\
  &\quad +
  (2 {\bf v}_{\mathrm{body}} \cdot {\bf v}_{\mathrm{flap}})
  C_{F,\mathrm{cp}}(\alpha). \label{eq:newQSM_tctd}
\end{align}
Unsurprisingly, each of these velocity terms varies in a different pattern in a wingbeat when the body velocity is nonzero (Fig. \ref{fig:60_new_mechanisms_explanation}(c)). Based on this decomposition, translational lift and drag (Eqs. \ref{eq:convQSM_simple_tc} and \ref{eq:convQSM_simple_td}) can be expressed as \cite{Dickson2004-jj}
\begin{align}
  \mathrm{d} F _{\mathrm{tc}} &+ \mathrm{d} F _{\mathrm{td}} \notag \\
  &\propto
  \| {\bf v}_{\mathrm{wing}} \|^2 
  (
    k_1 \sin (2 \alpha_x) \cos \alpha_x  + k_2 \sin^3 \alpha_x
  ) \notag \\
  &=
  (v_{\mathrm{w},x}^2 + v_{\mathrm{w},z}^2)
   (k_{1}' \sin \alpha_x + k_{2}' \sin^3 \alpha_x) \notag \\
  &= 
  ( 
    v_{\mathrm{w},x} v_{\mathrm{b},x} - v_{\mathrm{w},x} \omega_{\mathrm{w},z} p_{y} \notag \\
    & \quad ~ + v_{\mathrm{w},z} v_{\mathrm{b},z} + v_{\mathrm{w},z} \omega_{\mathrm{w},x} p_{y} - v_{\mathrm{w},z} \omega_{\mathrm{w},y} p_{x} 
  ) \notag \\
  &\quad ~ \times (k_{1}' \sin \alpha_x + k_{2}' \sin^3 \alpha_x) \\
  &= ( v_{\mathrm{b},x}^2 + v_{\mathrm{b},z}^2 ) \notag \\
  &\quad ~ \times (k_{1,\mathrm{tr}} \sin \alpha_x + k_{2,\mathrm{tr}} \sin^3 \alpha_x) \notag \\
  &\quad + ( \omega_{\mathrm{w},z}^2 p_{y}^2 + \omega_{\mathrm{w},x}^2 p_{y}^2 + \omega_{\mathrm{w},y}^2 p_{x}^2 -2 \omega_{\mathrm{w},x} \omega_{\mathrm{w},y} p_{y} p_{x}) \notag \\
  &\quad ~ \times (k_{1,\mathrm{fl}} \sin \alpha_x + k_{2,\mathrm{fl}} \sin^3 \alpha_x) \notag \\
  &\quad + ( 2 v_{\mathrm{b},x} \omega_{\mathrm{w},z} p_{y} + 2 v_{\mathrm{b},z} \omega_{\mathrm{w},x} p_{y} - 2 v_{\mathrm{b},z} \omega_{\mathrm{w},y} p_{x} ) \notag \\
  &\quad ~ \times (k_{1,\mathrm{cp}} \sin \alpha_x + k_{2,\mathrm{cp}} \sin^3 \alpha_x), 
\end{align}
where $[p_x, p_y, 0]$ represents the coordinates on the wing surface, and $k_x$ are the coefficients for each kinematic term in the conventional QSM, whereas those with subscripts (tr, fl, and cp for translation, flapping, and coupling) are the coefficients for the decomposed kinematic terms. In the equation above, we expressed the lift coefficient as being proportional to $\sin(2\alpha_x)$ for simplicity. Our algorithm identified some of the functions that appeared when the velocities were decomposed.
%
\begin{equation}
  v_{\mathrm{w},z} \omega_{\mathrm{w},x} \sin(\alpha_x), 
  v_{\mathrm{w},z} v_{\mathrm{b},z} \sin(\alpha_x), 
  v_{\mathrm{b},x} \omega_{\mathrm{w},z} \sin(\alpha_x). \notag
\end{equation}
This suggests that the speeds of the body and wingbeat must be decomposed, and different coefficients should be applied to each. Previous studies have indicated that the lift and drag coefficients are different when the wing translates and when it simply rotates \cite{Dickinson1999-ug,Lentink2009-lx}, which cannot be expressed by Eq. \ref{eq:convQSM_tctd} but by the linear combination of the velocity terms ($\|{\bf v}_{\mathrm{body}}\|^2$, $\|{\bf v}_{\mathrm{flap}}\|^2$, $2{\bf v}_{\mathrm{body}} \cdot {\bf v}_{\mathrm{flap}}$) in Eq. \ref{eq:newQSM_tctd}. The advance ratio effect expressed using Eq. \ref{eq:newQSM_tctd} has already been proposed previously \cite{Dickson2004-jj} but has not been considered in most subsequent studies \cite{Dickson2008-ex,Truong2011-wu,Armanini2016-qh,Cai2021-ve,Deng2006-rz} possibly because its relative importance was uncertain. Our algorithm rediscovered the effects overlooked by the community, highlighting the importance of improving the accuracy of QSM.

\begin{figure*}[tbp]
    \centering
    \includegraphics[width=1.0\textwidth]{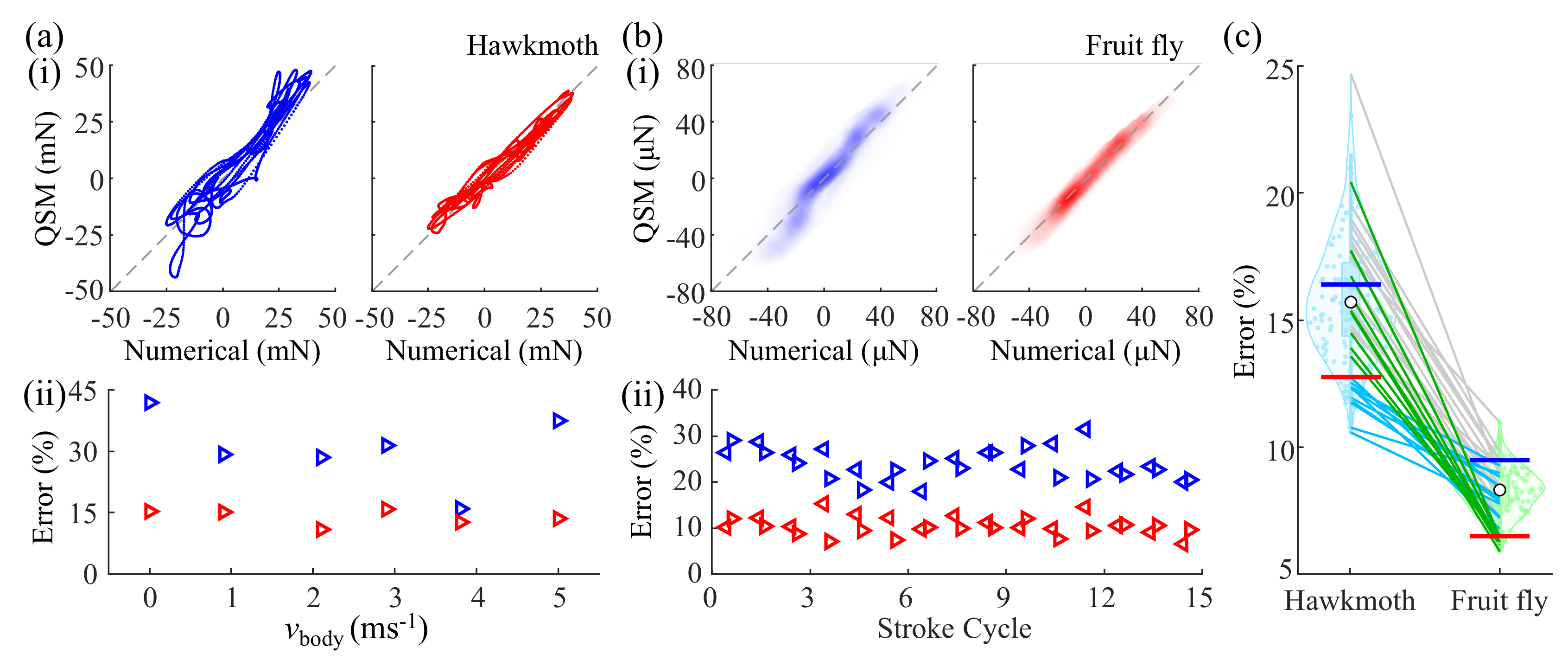}
    \caption{
      Improved prediction accuracy of aerodynamic forces using the data-driven QSM for (a) hawkmoths and (b) fruit flies. (i) Instantaneous aerodynamic forces predicted by the data-driven QSM (red) are closer to the CFD results (dashed lines) compared to those predicted by the conventional QSM (blue). (ii) Reduction of the RMSE in the estimate of aerodynamic forces by the data-driven QSM (red) relative to the conventional QSM (blue). 
      (c) Violin plots showing prediction errors when 24 functions are randomly selected from the candidate functions and added to the conventional QSM (blue lines). 
      The lines connect the errors from the same random models applied to hawkmoth and fruit fly cases. 
      While some randomly generated models yield errors smaller than that of the data-driven QSM (red line) in either the hawkmoth case (cyan) or the fruit fly case (green), they tend to exhibit large errors when applied to another case (green and cyan lines). White circles represent the median of the errors.
    }
    \label{fig:70_imporved_acuaracy}
\end{figure*}

The second mechanism is the effect of the spanwise kinematic velocity (from the wing base to the wing tip, or vice versa). This mechanism is similar to the translational lift, drag, and rotational circulation mechanisms described in conventional QSMs; however, it is caused by the air velocity of the wing along the wingspan, namely the spanwise kinematic velocity (Figs. \ref{fig:60_new_mechanisms_explanation}(b) and (d)). The conventional QSMs for free flight consider air velocity of the chordwise blade (gray cross-section in Fig. \ref{fig:60_new_mechanisms_explanation}(b)) in chordwise direction, $v_{\mathrm{air},x}$, and in the direction normal to the wing surface, $v_{\mathrm{air},z}$, but not air velocity in the spanwise direction, $v_{\mathrm{air},y}$. By considering the spanwise blade (black cross-section in Fig. \ref{fig:60_new_mechanisms_explanation}(b)), the translational lift $\mathrm{d}F_{\mathrm{tc,span}}$ and drag $\mathrm{d}F_{\mathrm{td,span}}$ and the rotational circulatory force $\mathrm{d}F_{\mathrm{rc,span}}$ due to $v_{\mathrm{air},y}$ (Fig. \ref{fig:60_new_mechanisms_explanation}(d)) can be expressed as follows:
%
\begin{align}
  &\mathrm{d} F _{\mathrm{tc,span}} + \mathrm{d} F _{\mathrm{td,span}} \notag \\
  &\propto
  v_{\mathrm{w},yz}^2
  (
    k_1 \sin (2 \alpha_y) \cos \alpha_y  + k_2 \sin^3 \alpha_y
  ) \notag \\
  &=
  (v_{\mathrm{w},y}^2 + v_{\mathrm{w},z}^2)
   (k_{1}' \sin \alpha_y + k_{2}' \sin^3 \alpha_y) \notag \\
  &=
  ( 
    v_{\mathrm{w},y} v_{\mathrm{b},y} + v_{\mathrm{w},y} \omega_{\mathrm{w},z} p_{x} \notag \\
    & \quad ~ + v_{\mathrm{w},z} v_{\mathrm{b},z} + v_{\mathrm{w},z} \omega_{\mathrm{w},x} p_{y} - v_{\mathrm{w},z} \omega_{\mathrm{w},y} p_{x} 
  ) \notag \\ 
  & \quad ~ \times (k_{1}' \sin \alpha_y + k_{2}' \sin^3 \alpha_y) \\ 
  &= ( v_{\mathrm{b},y}^2 + v_{\mathrm{b},z}^2 ) \notag \\
  & \quad ~ \times (k_{1,\mathrm{tr}} \sin \alpha_y + k_{2,\mathrm{tr}} \sin^3 \alpha_y) \notag \\
  & \quad + ( \omega_{\mathrm{w},z}^2 p_{x}^2 + \omega_{\mathrm{w},x}^2 p_{y}^2 + \omega_{\mathrm{w},y}^2 p_{x}^2 -2 \omega_{\mathrm{w},x} \omega_{\mathrm{w},y} p_{x} p_{y}) \notag \\
  & \quad ~ \times (k_{1,\mathrm{fl}} \sin \alpha_y + k_{2,\mathrm{fl}} \sin^3 \alpha_y) \notag \\
  & \quad + ( 2 v_{\mathrm{b},y} \omega_{\mathrm{w},z} p_{x} + 2 v_{\mathrm{b},z} \omega_{\mathrm{w},x} p_{y} - 2 v_{\mathrm{b},z} \omega_{\mathrm{w},y} p_{x} ) \notag \\
  & \quad ~ \times (k_{1,\mathrm{cp}} \sin \alpha_y + k_{2,\mathrm{cp}} \sin^3 \alpha_y), 
\end{align}
\begin{align}
  &\mathrm{d} F _{\mathrm{rc,span}} 
  \propto
  v_{\mathrm{w},y} \omega_{\mathrm{w},x} = (v_{\mathrm{b},y} + \omega_{\mathrm{w},z} p_{x}) \omega_{\mathrm{w},x}.
\end{align}
Our algorithm identified some of the functions that represent the contribution of the spanwise kinematic velocity as follows:
%
\begin{gather}
  v_{\mathrm{w},z} \omega_{\mathrm{w},y} \sin(\alpha_y), 
  v_{\mathrm{b},z}^2 \sin(\alpha_y), 
  v_{\mathrm{b},z} \omega_{\mathrm{w},y} \sin^3(\alpha_y), \notag \\
  v_{\mathrm{w},y} \omega_{\mathrm{w},x},
  v_{\mathrm{b},y} \omega_{\mathrm{w},x}. \notag
\end{gather}
While the effect of the spanwise kinematic velocity on flapping-wing aerodynamics has been verified qualitatively \cite{Han2019-vn}, it has not been formulated previously. Our algorithm highlights the importance of this effect and provides mathematical formulations suggesting that the concept of a spanwise blade element would likely not have emerged without this data-driven formulation. In hawkmoth forward flight ($v_{\mathrm{body}} = 5.0 \mathrm{ms}^{-1}$), for example, the translational force by the spanwise kinematics contributes positively (from the ventral to the dorsal surface of the wing) throughout the wingbeat (light green in Figs. \ref{fig:60_new_mechanisms_explanation}(d) and (e)). The rotational circulatory lift by the spanwise kinematics contributes positively at the beginning of the downstroke and negatively (from the dorsal to the ventral surface of the wing) during the upstroke in the hawkmoth forward flight, whereas the contribution is relatively small at the supination (dark green in Figs. \ref{fig:60_new_mechanisms_explanation}(d) and (e)). This asymmetry is possibly because the spanwise kinematic velocity (from the wing tip to the wing base) is relatively high during pronation, whereas the wing is supinated at the side of the body in this case, and the opposing spanwise kinematic velocity (owing to body motion) is lower.

The third mechanism, the rotational Wagner effect, is an extension of the Wagner effect proposed by van Veen et al. \cite{van-Veen2023-dq} to the rotational circulation. They numerically demonstrated that in a flapping flight, the development and decay of the circulation generated by translation are delayed by the acceleration and deceleration of the wings, respectively. Using a similar theory for rotational circulation, the decay and development of rotational circulation can be delayed by the acceleration and deceleration of the wings, respectively (Fig. \ref{fig:60_new_mechanisms_explanation}(f); see Appendix \ref{subsec:App_formulationWagner} for the derivation). This effect is formulated as follows:
\begin{align}
  \mathrm{d} F_{\mathrm{rc,Wg}} \propto
  \left(
    \omega_{\mathrm{w},y}(t_{\mathrm{eq}}) - \omega_{\mathrm{w},y}
  \right) \mathrm{sign}(a_{\mathrm{w},x}) \sqrt{|a_{\mathrm{w},x}|}.
\end{align}
One of the terms identified by our algorithm is part of the formula as follows:
%
\begin{align}
  \omega_{\mathrm{w},y} \mathrm{sign}(a_{\mathrm{w},x}) \sqrt{|a_{\mathrm{w},x}|}. \notag
\end{align}
To the best of our knowledge, this mechanism has not been identified explicitly in previous studies. The selection of functions using our algorithm extends the translational Wagner effect discussed in a previous study \cite{van-Veen2023-dq} to its rotational counterpart. This mechanism contributes noticeably to the flight of the fruit fly (i3 in Figs. \ref{fig:40_kinematics_data_HMFF}(c) and (d)), particularly at the early stage of each stroke (Fig. \ref{fig:60_new_mechanisms_explanation}(g)). The rotational Wagner effect delays the decay of the rotational circulation; therefore, when the wing rotation (pitching down for the stroke) slows down while the wing is accelerating (vertical broken lines in Fig. \ref{fig:60_new_mechanisms_explanation}(g)), it counteracts the reduction in force owing to the rotational circulation, resulting in an increase in the force generated by this mechanism.

\begin{figure*}[tbp]
    \centering
    \includegraphics[width=1.0\textwidth]{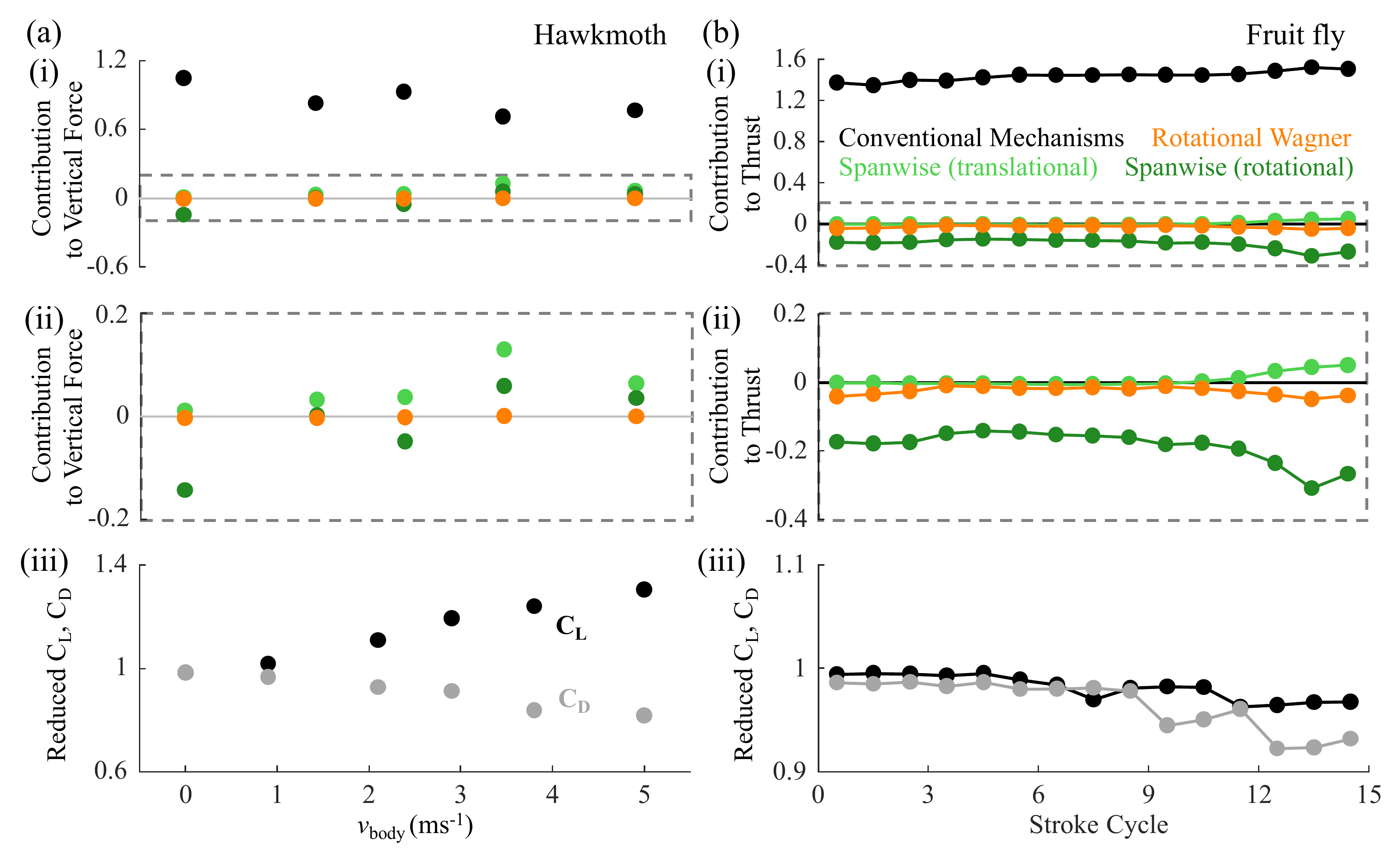}
    \caption{
      Contributions of newly identified mechanisms to aerodynamic force generation in the (a) vertical force of hawkmoths and (b) thrust of fruit flies. (i) Effects of spanwise translational and rotational forces and the rotational Wagner effect. (ii) Magnified view of (i). (iii) Variations in the reduced lift and drag coefficients illustrating the effect of the advance ratio.
    }
    \label{fig:80_new_mechanisms_contribution}
\end{figure*}

\begin{figure*}[tbp]
    \centering
    \includegraphics[width=1.0\textwidth]{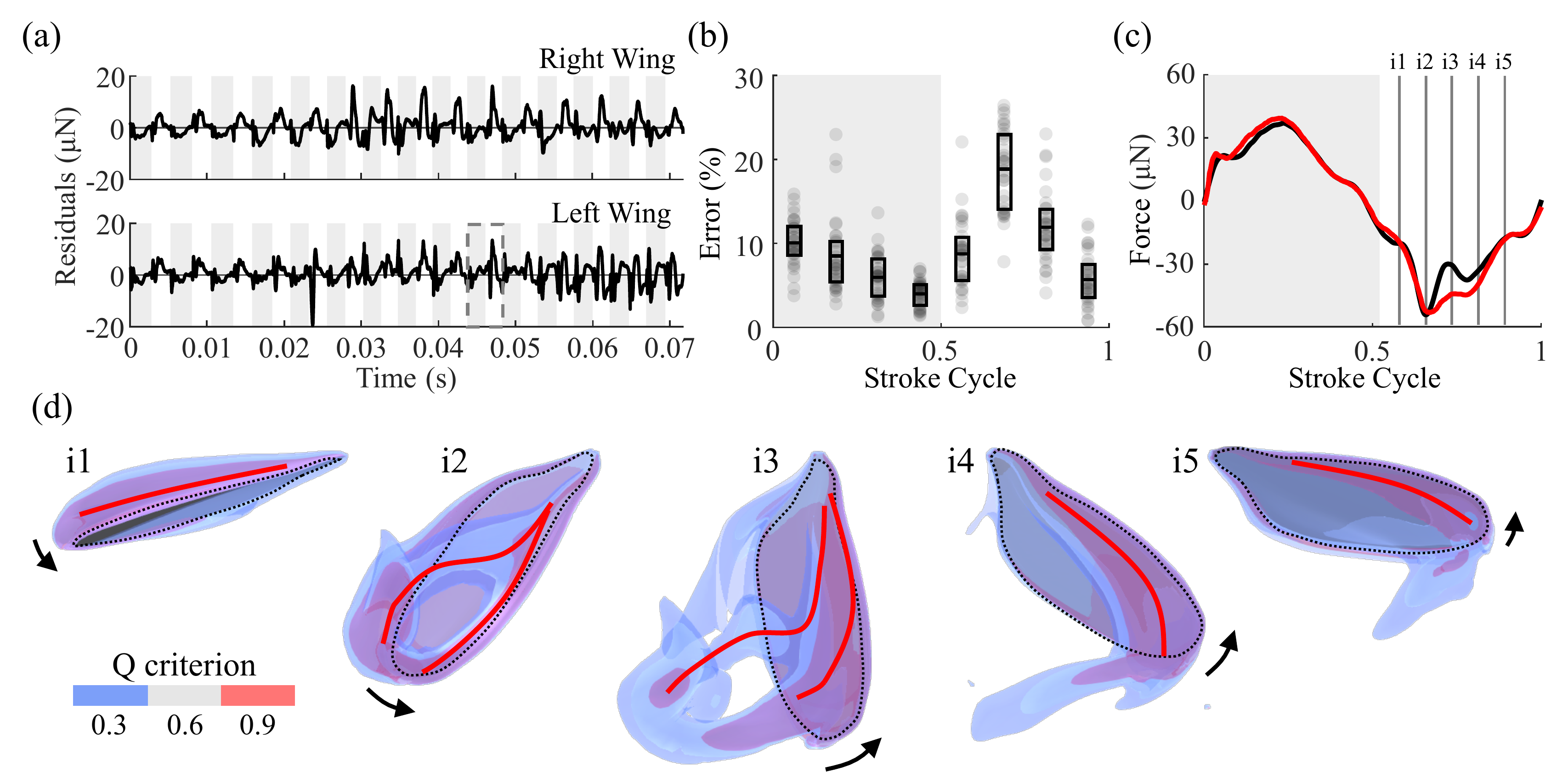}
    \caption{
      Residual errors remaining in the data-driven QSM. (a) Time series of residuals for the aerodynamic forces generated by the right wing (top) and left wing (bottom) of fruit flies. (b) Mean (black line) and standard deviation (box) of the error over a wingbeat cycle. The raw data are shown as gray circles. (c) Time series of aerodynamic forces computed by CFD (black) and the data-driven QSM (red) in the dashed box of (a). (d) Iso-surfaces of the Q-criterion around the left wing during upstroke for time instances in the range of i1 to i5 in (c). A breakdown of the vortex structure is observed at i3, where the error is large. The gray shaded areas in (a,b,c) represent the downstroke.
    }
    \label{fig:90_error_DDQSM}
\end{figure*}

%
\subsection{Improvement in the accuracy of force estimate \label{subsec:improvementInAccuracy}}

By incorporating the three mechanisms described in Sec. \ref{subsec:identifiedMechanisms}, a new data-driven QSM for free flight was constructed (see Appendix \ref{subsec:App_dataDrivenQsm} for a full description). To confirm its utility, we used data-driven QSM to estimate the aerodynamic forces generated by the flight of the hawkmoth and fruit fly, including hovering, forward, and maneuvering flights (Figs. \ref{fig:40_kinematics_data_HMFF}, \ref{fig:70_imporved_acuaracy}(a), and \ref{fig:70_imporved_acuaracy}(b)). The model coefficients were determined using the Levenberg-Marquardt algorithm (using the method expressed in Sec. \ref{subsubsec:modelDiscovery}; see Appendix \ref{sec:A2_App_convmodel} for details), and the resulting values were found to be physically reasonable when compared with previous experimental studies (see Appendix \ref{subsec:App_coeffecients} for details).

The data-driven QSM showed a substantial reduction in the estimation error of the instantaneous aerodynamic forces for all flight modes compared to the conventional QSM (Figs. \ref{fig:70_imporved_acuaracy}(a) and (b)). The conventional QSM overestimates large instantaneous aerodynamic forces (blue in Figs. \ref{fig:70_imporved_acuaracy}(a)i and (b)i), and the RMSE of the estimated instantaneous aerodynamic forces varies significantly depending on the flight mode (blue in Figs. \ref{fig:70_imporved_acuaracy}(a)ii and (b)ii). However, the consideration of the three novel mechanisms corrects for errors in larger instantaneous aerodynamic forces (red in Figs. \ref{fig:70_imporved_acuaracy}(a)i and (b)i). As a result, the data-driven QSM greatly reduces the variation in the estimation error of the aerodynamic force across flight modes (11.0--16.0\% for hawkmoth, 6.7--15.4\% for fruit flies; red in Figs. \ref{fig:70_imporved_acuaracy}(a)ii and (b)ii). These results demonstrate the versatility of the data-driven QSM in being able to accurately estimate the aerodynamic forces over a wide range of flight modes (0.0--1.1 of the advance ratio) and $Re$ (190--290 and 6100--11000 based on the mean wingtip velocity). This improvement is achieved by considering the three new mechanisms described above, which allow a better representation of the large variations in aerodynamic forces owing to complex wing kinematics, including body motion.

Our data-driven QSM is a linear combination of 34 terms, which seems redundant compared to the conventional QSM (10 terms). However, the reduction in error in the above wide regime was achieved not by redundancy alone but by the selection of the appropriate functions. To demonstrate this, we examined the error of 1,000 different models with 24 randomly selected candidate terms in addition to the 10 terms of the conventional QSM (Eq. \ref{eq:conventional_QSM}). Compared with the conventional QSM (blue line in Fig. \ref{fig:70_imporved_acuaracy}(c)), several random models reduce the error by increasing the number of terms. In fact, there are random models with errors smaller than our data-driven QSM (red line in Fig. \ref{fig:70_imporved_acuaracy}(c)) for both the hawkmoth and fruit fly models. However, the random model that reduced the error in one case exhibited a higher error than that in the data-driven QSM in another case (see the cyan and green lines in Fig. \ref{fig:70_imporved_acuaracy}(c)). This wide range of applicability to various flight modes and Reynolds numbers was observed because the functions were not selected randomly; rather, they were appropriately selected based on the aerodynamics, from among the important functions. The model constructed in this study was considered highly accurate and versatile, even considering its complexity.

The improvement in the accuracy of QSM in hawkmoth forward flight was primarily attributed to the incorporation of the effects of advance ratio and spanwise kinematic velocity. To quantitatively evaluate the contributions of the spanwise kinematic velocity and rotational Wagner effect, we calculated the contribution of each mechanism by dividing the cycle average of the vertical aerodynamic force from each mechanism by that of the total vertical aerodynamic force (Fig. \ref{fig:80_new_mechanisms_contribution}(a)). The contributions from the conventional mechanisms were close to one, explaining a substantial portion of the aerodynamic forces (Fig. \ref{fig:80_new_mechanisms_contribution}(a)i). However, the contribution of spanwise kinematics was also significant, accounting for more than 10\% of the aerodynamic force at several flight speeds (Fig. \ref{fig:80_new_mechanisms_contribution}(a)ii). The translational force generated by the spanwise kinematics increases the vertical force at a higher flight speed. The rotational force generated by the spanwise kinematics reduced the vertical force during hovering. As the flight speed increased, this contribution diminished or even reversed, increasing the vertical force. These results imply that even in forward flight, the spanwise component of the wing velocity contributes considerably to aerodynamic force generation. The rotational Wagner effect is relatively small in the forward flight of the hawkmoth (Fig. \ref{fig:80_new_mechanisms_contribution}(a)ii).

The influence of the advance ratio can be assessed using the reduced coefficients (i.e., the translational lift and drag terms in the data-driven QSM divided by the corresponding conventional kinematic terms) as follows:
\allowdisplaybreaks[0]
\begin{align}
  \hat{C}_{F}(\alpha) &=
  \left(
    \| {\bf v}_{\mathrm{body}} \|^2 C_{F,\mathrm{bd}}(\alpha) 
    + \| {\bf v}_{\mathrm{flap}} \|^2 C_{F,\mathrm{fl}}(\alpha) 
    \right. \notag \\
    &\quad \left.
     + 2 ({\bf v}_{\mathrm{body}} \cdot {\bf v}_{\mathrm{flap}}) C_{F,\mathrm{cp}}(\alpha)
    \right) / \| {\bf v}_{\mathrm{body}} + {\bf v}_{\mathrm{flap}} \|^2. 
\end{align}
\allowdisplaybreaks[4]
Note that, in Figs. \ref{fig:80_new_mechanisms_contribution}(a)iii and (b)iii, the coefficients are further divided by the coefficients at hovering to emphasize their variation. The reduced coefficients will always be constant in the conventional QSM; however, in the data-driven QSM, these coefficients can be varied to consider the effect of the advance ratio. In the hawkmoth forward flight, the reduced lift and drag coefficients monotonically increase or decrease with the flight speed (Fig. \ref{fig:80_new_mechanisms_contribution}(a)iii), implying a significant contribution of the advance ratio effect to the fast forward flight.

In the maneuvering flight of a fruit fly, which transitions from a slow flight through acceleration and turning to a relatively fast forward flight (Figs. \ref{fig:40_kinematics_data_HMFF}(c) and (d)), the newly identified mechanisms contribute significantly and cannot be neglected (Fig. \ref{fig:80_new_mechanisms_contribution}(b)). The effect of the spanwise kinematic velocity increases as the flight speed increases, with the translational force by the spanwise kinematics acting to enhance the thrust at higher speeds and the rotational force by the spanwise kinematics acting to reduce the thrust (Fig. \ref{fig:80_new_mechanisms_contribution}(b)ii). The rotational Wagner effect made a relatively large contribution during steady flight at three to four wingbeats at the beginning and end of the sequence (Fig. \ref{fig:80_new_mechanisms_contribution}(b)ii). The reduced coefficients for the translational lift and drag decreased from turning to fast flight (Fig. \ref{fig:80_new_mechanisms_contribution}(b)iii), similar to the forward flight of the hawkmoth (Fig. \ref{fig:80_new_mechanisms_contribution}(a)iii). These results show that the fruit fly maneuvers by adjusting the aerodynamic mechanisms associated with complex movements of the body.

Thus far, we have shown that introducing three mechanisms improved QSM performance and provided insights into insect flight aerodynamics; however, estimation errors remain. The data-driven QSM error for fruit flies is high mainly during upstroke (Figs. \ref{fig:90_error_DDQSM}(a) and (b)). Differences between the data-driven QSM and CFD forces are large (Fig. \ref{fig:90_error_DDQSM}(c) i2--i3) during vortex breakdown in upstroke (Fig. \ref{fig:90_error_DDQSM}(d)) and small (Fig. \ref{fig:90_error_DDQSM}(c) i4--i5) when the leading-edge vortex is stable. This corresponds to reduced aerodynamic forces when the leading-edge vortex breaks down \cite{Sane2003-tm,Ellington1996-ze,Harbig2014-sd}. The results show that QSM cannot account for aerodynamic force variations stemming from flow instability (Fig. \ref{fig:90_error_DDQSM}(c)).

%
\section{\label{sec:4_discussion}DISCUSSION}
In this study, we explored a mathematical representation of aerodynamic forces generated by flapping wings in general, not limited to hovering flight, using a machine learning algorithm. Despite its utility and benefits, to date, most studies on QSM have been limited to hovering \cite{Sane2001-gl,Berman2007-tg,Nakata2015-fr,Lee2016-hh,Weis-Fogh1973-km,Sane2002-ch,Whitney2010-fr} or forward flight \cite{Cai2021-ve,Dickson2004-jj,Dudley1990-dr}; studies on extending the model for forward and free-flight modes are limited by the challenges associated with the free-flight mode---a three-dimensional, complex phenomenon involving the flapping of wings as well as translation and rotation of the body, whose evaluation via intuition or extended conventional revolving wing models is arduous. Data-driven approaches that utilize data to identify orders in complex systems (such that important elements can be identified from a large number of candidates) are well suited to this task. By leveraging the characteristics of the data-driven approach, this study challenged the limitations of the QSM and achieved a drastic improvement in accuracy (Fig. \ref{fig:70_imporved_acuaracy}), representing a step further away from the observation of correlations in simplified systems.

The three new mechanisms are physically reliable because they are described by physically interpretable formulas that have been underpinned by previous studies. Data-driven approaches may produce mathematical formulas that are not physically meaningful depending on the training data \cite{Champion2019-wy,Song2024-hz}. However, the mechanisms we added after careful selection and investigation were effects that previous studies had indicated a correlation. The advance ratio effect has been explicitly highlighted in \cite{Lentink2009-lx,Dickson2004-jj}. The effect of spanwise kinematic velocity was indirectly noted as the lateral flow effect \cite{Han2019-vn}. The rotational Wagner effect is an extension of the translational Wagner effect associated with acceleration and deceleration \cite{van-Veen2023-dq}. The significance of these mechanisms (relative to conventional mechanisms) was not certain and has been truncated or not formulated in previous studies because of insufficient data. Our data-driven approach highlighted their importance, enabled their formulation, and greatly improved the accuracy of QSM across a range of $Re$ values (Fig. \ref{fig:70_imporved_acuaracy}(c)).

The present data-driven QSMs enable a detailed investigation of the effects of wing kinematics on the performance of flyers, such as efficiency, stability, and maneuverability. Understanding the mechanism by which and when insects adjust their wings is crucial from a neurophysiological perspective \cite{Melis2024-jk,Lehmann2017-ef,Dickinson2016-zz}. The effects of fine-tuning of the wing motion have been investigated with precision using time-consuming approaches such as computational \cite{Haque2023-ur,Ramamurti2007-vo} or robotic simulations \cite{Muijres2014-hx,Fry2003-xr,Muijres2015-ay}; however, with an accurate QSM, the aerodynamic performance of wing motion can be investigated quickly. By breaking down the aerodynamic forces into components, as shown in Sec. \ref{subsec:improvementInAccuracy} (Fig. \ref{fig:80_new_mechanisms_contribution}), and clarifying the relative importance of the mechanisms for the flight mode under study, the functional roles of specific wing kinematic adjustments can be interpreted in detail. This step can provide insights into the aerodynamic consequences of the measured wing motions, offering a basis for generating testable hypotheses about control strategies rather than merely describing kinematics. Although the data-driven QSM requires training data to acquire its coefficients, similar to interpolation using neural networks \cite{Corban2023-aw,Hu2024-oc,Lee2019-ra,Lino2023-is}, these potential benefits can only be achieved by the mathematical formulation of aerodynamic forces, rather than from the simulation or interpolation of phenomena.

The present data-driven QSM, built from over 5,000 candidate terms, cannot fully reproduce aerodynamic forces based on high-fidelity CFD (Fig. \ref{fig:90_error_DDQSM}), suggesting future research directions. To understand vortex breakdown, which causes large errors (Fig. \ref{fig:90_error_DDQSM}), understanding at what kinematics and histories it occurs and to what extent the lift is reduced is necessary. The wake-capture effect \cite{Dickinson1999-ug,Xuan2020-ca,Poletti2024-fm,Sane2003-tm,Birch2003-cd}, not considered here, involves numerous physical variables, whose quasi-steady modeling is challenging. These fluid phenomena, such as vortex breakdown and wake capture, which are strongly influenced by fluid motion, may require stochastic treatment \cite{Boninsegna2017-vg,Callaham2021-vl}.

Diversity of insect morphology and behavior presents further challenges to aerodynamic modeling. 
Although the two species considered in this study, hawkmoths and fruit flies, are functionally two-winged flyers \cite{Willmott1997-ow,Fry2003-xr}, insects in general exhibit a wide range of morphological and behavioral traits that can result in more complex aerodynamic phenomena. For example, beetles and dragonflies fly using four wings \cite{Le2013-zl,Wang2005-by}. The wing-wing interaction between the anterior and posterior wings of these species influences the generation of aerodynamic forces \cite{Le2013-zl,Wang2005-by}. Thin insect wings deform passively under aerodynamic and inertial forces, and their morphology is dynamically adjusted, resulting in significantly altered performances \cite{Nakata2012-og,Hsu2024-uw,Young2009-ep}. In addition, aerodynamic forces are strongly variable when insects take off because of ground effects \cite{Kolomenskiy2016-vv}. Using the current approach, these additional effects can be mathematically represented by extending the training data and adding additional parameters. However, such expansions would dramatically increase the size of the dataset and the number of candidate functions. Novel efficient strategies will be required to manage this complexity and maintain tractability. 

This model is also useful for robot design. Revolving wings are more efficient when hovering, at least for Reynolds numbers above $10^2$ \cite{Bayiz2018-im,Zheng2013-is}. In terms of efficiency, the advantages of flapping wings may be exploited in forward flight, where the relative wind speed, and hence the lift due to body movement, can be utilized. For example, intermittent flights utilizing gliding with the wings stopped between flapping flights, although not performed by the target insects in this study, would be useful for improving efficiency \cite{Tobalske2010-vs,Pennycuick2008-xq,Betts1988-ng,Paoletti2014-bn}. Flapping wings may have an advantage in flight maneuvering owing to abrupt force changes, because their degrees of freedom exceed those of revolving wings, and they allow fine-tuning of wing movements \cite{Muijres2014-hx,Muijres2015-ay}. The movement and control of wings to optimize these benefits may necessitate intricate mechanism design, akin to the complex musculoskeletal systems and joints that insects have developed through evolutionary processes \cite{Melis2024-jk,Deora2017-mw}. The data-driven QSM developed in this study could be a crucial aid in this design.

%
\section{\label{sec:5_conclusion}CONCLUSION}
In this study, we aimed to enhance the accuracy of quasi-steady models (QSMs), which are useful for rapid prediction of aerodynamic forces generated by flapping wings in insect flight. Using a data-driven approach, we identified previously overlooked but essential mechanisms that significantly contribute to force generation. As a result, we successfully formulated three mechanisms that had been absent from existing QSMs: the advance ratio effect, effect of spanwise kinematic velocity, and rotational Wagner effect. By incorporating these mechanisms, the data-driven QSM achieved substantial improvement in accuracy across hovering, forward, and maneuvering flight for a wide range of Reynolds number ($10^2$--$10^4$). The resulting model enables fast aerodynamic performance evaluation and functions as a valuable tool for studying insect flight dynamics from physical, biological, and engineering perspectives.

\begin{acknowledgments}
This work was supported by the JST BOOST, Japan, Grant Number JPMJBS2413 to YK, and JSPS KAKENHI Grant Numbers JP24K00829 and JP24K21277 to TN.
\end{acknowledgments}

\appendix

\begin{table}[tbp]
  \caption{\label{tab:2_nomancleture}%
  List of notations. Angular velocities, angular accelerations, translational velocities, and translational accelerations are defined relative to the global frame. The vectors are decomposed into the wing-fixed frame unless otherwise noted. The elements of vectors are written with the suffix $x$, $y$ or $z$. ${\bf q}$ denotes arbitrary vectors.
  }
  \begin{ruledtabular}
    \begin{tabular}{cc}
      $x_{\mathrm{g}}, y_{\mathrm{g}}, z_{\mathrm{g}}$ & Axes of global frame \\
      $x_{\mathrm{sp}}, y_{\mathrm{sp}}, z_{\mathrm{sp}}$ & Axes of stroke-plane-fixed frame \\
      $x_{\mathrm{w}}, y_{\mathrm{w}}, z_{\mathrm{w}}$ & Axes of wing-fixed frame \\
      $\theta_{\mathrm{pos}}$ & Positional angle \\
      $\theta_{\mathrm{ele}}$ & Elevation angle \\
      $\theta_{\mathrm{fea}}$ & Feathering angle \\
      $\theta_{\mathrm{SP}}$ & Stroke plane angle \\
      ${\bf R}_{\mathrm{g,w}}$ & Transformation matrix from global frame \\
       &  to wing-fixed frame \\
      $\bm{\omega}_{\mathrm{w}}$ & Angular velocity of wing \\
      $\bm{\omega}_{\mathrm{b}}$ & Angular velocity of body \\
      $\bm{\psi}_{\mathrm{w}}$ & Angular acceleration of wing \\
      ${\bf v}_{\mathrm{b}}$ & Velocity of center of mass of the body \\
      ${\bf v}_{\mathrm{w}}$ & Velocity of a point on the wing \\
      ${\bf a}_{\mathrm{w}}$ & Acceleration of a point on the wing \\
      ${\bf p}$ & Position of a point on the wing \\
      $\alpha_x$ & Angle of attack in chordwise direction \\
      $\alpha_y$ & Angle of attack in spanwise direction \\
      $t$ & Time \\
      $g$ & Gravity acceleration \\
      $\rho$ & Density of air \\
      $\nu$ & Kinetic viscosity of air \\
      $c$ & Chord length \\
      $c_{\mathrm{m}}$ & Mean chord length \\
      $s$ & Span length \\
      $R$ & Wing length \\
      $Re$ & Reynolds number \\
      $q_x$ & $x$ component of vector {\bf q} \\
      $q_y$ & $y$ component of vector {\bf q} \\
      $q_z$ & $z$ component of vector {\bf q} \\
      $q_{xz}$ & $\sqrt{q_x^2 + q_z^2}$ \\
      $q_{yz}$ & $\sqrt{q_y^2 + q_z^2}$ \\
    \end{tabular}
  \end{ruledtabular}
\end{table}

%
\section{\label{sec:A1_App_nomenclature}NOMENCLATURE}

The symbols used in this study are listed in Table \ref{tab:2_nomancleture}. The $x$, $y$, and $z$ components of the vectors are expressed in a wing-fixed frame, unless otherwise noted.

%
\section{\label{sec:A2_App_convmodel}CONVENTIONAL QSM}

As described in Sec. \ref{subsubsec:conventionalQSM}, the conventional QSM is formulated based on the BEM and is composed of translational lift and drag, rotational lift and drag, and the force due to the added mass (Eqs. \ref{eq:convQSM_simple_tc}--\ref{eq:convQSM_simple_am}) with the Wagner effect and Reynolds number effects. Because integration over the blade is computationally expensive, we developed a method to avoid integration while maintaining model accuracy. For instance, in the case of a translational lift, the integration is approximated as
\begin{align}
  F_{\mathrm{tc}} &= \int \mathrm{d} F_{\mathrm{tc}} \notag \\
  &=
  \frac{1}{2} \rho 
  \int
    v_{\mathrm{w},xz}^2({\bf p})
    C_{L}(\alpha_x({\bf p})) \cos(\alpha_x({\bf p})) c \mathrm{d}y \label{eq:analyticalFtc} \\
  &\fallingdotseq
  \frac{1}{2} \rho 
  v_{\mathrm{w},xz}^2(\widehat{\bf p})
  C_{L}(\alpha_{x}(\widehat{\bf p}))
  \cos({\alpha_{x}(\widehat{\bf p})})
  \int c \mathrm{d}y, \label{eq:approxFtc}
\end{align}
where we rewrote the velocity and angle of attack as $v_{\mathrm{w},xz}({\bf p})$, $\alpha_x({\bf p})$ to emphasize that they depend on the point on the wing ${\bf p} = [p_x, p_y,0]$. $\widehat{{\bf p}}$ is a representative point for calculating the velocities and angle of attack and is determined by the wing morphology. To derive $\widehat{{\bf p}}$, we randomly sampled a large set of body velocities and angular velocities ($N = 10^7$), computed $v_{\mathrm{w},xz}({\bf p})$ and $\alpha_x({\bf p})$ for each point on the wing surface ${\bf p}$, and derived $\widehat{{\bf p}}$ that maximizes the correlation between Eqs. \ref{eq:analyticalFtc} and \ref{eq:approxFtc}. This procedure was performed for all mechanisms (see supplementary data).  Because the representative points are determined by wing morphology, they need not be recalculated during optimization or machine learning applications once computed. Eventually, the formulations of the mechanisms are given by:
\begin{align}
  F_{\mathrm{tc}} 
  &\fallingdotseq
  \frac{1}{2} \rho 
  v_{\mathrm{w},xz}^2(\widehat{\bf p})
  C_{L}(\alpha_{x}(\widehat{\bf p}))
  \cos({\alpha_{x}(\widehat{\bf p})})
  S_{\mathrm{w}},  \\
  F_{\mathrm{td}} 
  &\fallingdotseq
  \frac{1}{2} \rho 
  v_{\mathrm{w},xz}^2(\widehat{\bf p})
  C_{D}(\alpha_{x}(\widehat{\bf p}))
  \sin({\alpha_{x}(\widehat{\bf p})})
  S_{\mathrm{w}},  \\
  %
  F_{\mathrm{rc}}
  &\fallingdotseq
  - \rho v_{\mathrm{w},x}(\widehat{\bf p}) \beta_{\mathrm{rc}} \omega_{\mathrm{w},y} S_{c},  \\
  F_{\mathrm{rd}}
  &\fallingdotseq
  \frac{1}{2} \rho 
  \omega_{\mathrm{w},y} | \omega_{\mathrm{w},y} | 
  \beta_{\mathrm{rd}} S_{x|x|},  \\
  F_{\mathrm{AM}}
  &\fallingdotseq
  - \beta_{\mathrm{am,trans}} \frac{\pi}{4} \rho a_{\mathrm{w},z}(\widehat{\bf p}) S_{c} \notag \\
  &\quad +
  \beta_{\mathrm{am,rot}} \frac{\pi}{4} \rho \psi_{\mathrm{w},y} S_{x_m c}, 
\end{align}
where
\begin{align}
  S_{\mathrm{w}} &= 
  \iint_{\mathrm{wing}} ~ \mathrm{d}x \mathrm{d}y, \\
  S_{c} &= 
  \iint_{\mathrm{wing}} c~ \mathrm{d}x \mathrm{d}y, \\
  S_{x|x|}  &= 
  \iint_{\mathrm{wing}} x|x|~ \mathrm{d}x \mathrm{d}y, \\
  S_{x_m c} &= 
  \iint_{\mathrm{wing}} \frac{1}{2}(x_{\mathrm{TE}} + x_{\mathrm{LE}}) c  ~ \mathrm{d}x \mathrm{d}y.
\end{align}
The Wagner effect is defined as the delay in the development and decay of translational circulation (i.e., circulation produced by wing translation) owing to the unsteady motion of the wing \cite{Ellington1984-li} or, for flapping wings, the acceleration and deceleration \cite{van-Veen2023-dq}. We utilized the formulation by van Veen et al. \cite{van-Veen2023-dq} with some modifications to account for unsteady acceleration or deceleration, as follows:
\begin{align}
  F_{\mathrm{tc}\mathrm{,Wg}}
  &=
  \frac{1}{2} \rho 
  \left(
      v_{\mathrm{w},xz}(t_{\mathrm{eq}}) C_{L}(\alpha_x (t_{\mathrm{eq}})) -
      v_{\mathrm{w},xz} C_{L}(\alpha_x)
  \right) \notag \\
  &\quad \times
  \mathrm{sign}(a_{\mathrm{w},x}) \sqrt{c|a_{\mathrm{w},x}|} 
  \cos({\alpha_{x}})
  \int c \mathrm{d}y,  \\
  F_{\mathrm{td}\mathrm{,Wg}}
  &=
  \frac{1}{2} \rho 
  \left(
      v_{\mathrm{w},xz}(t_{\mathrm{eq}}) C_{D}(\alpha_x (t_{\mathrm{eq}})) -
      v_{\mathrm{w},xz} C_{D}(\alpha_x)
  \right) \notag \\
  &\quad \times
  \mathrm{sign}(a_{\mathrm{w},x}) \sqrt{c|a_{\mathrm{w},x}|} 
  \sin({\alpha_{x}})
  \int c \mathrm{d}y,  
\end{align}
where $F_{\mathrm{tc,wg}}$ and $F_{\mathrm{td,wg}}$ are the translational lift and drag terms, respectively. Further details are provided in Appendix \ref{subsec:App_formulationWagner}.

The coefficients of aerodynamic mechanisms are influenced by the Reynolds number. This Reynolds number effect is particularly important for flapping wings because the Reynolds number varies throughout the wingbeat with changes in the wing speed on a flapping wing. Based on previous experimental studies \cite{Lee2016-hh,Lentink2009-lx}, the Reynolds number effects $f_{Re,\mathrm{LD}}$ and $f_{Re,\mathrm{am}}$ for the lift, drag, and force due to added mass were formulated as
\begin{align}
  f_{Re,\mathrm{LD}} &= \exp(\gamma_{1} Re^{\gamma_{2}}),  \\
  f_{Re,\mathrm{am}} &= (1 + \gamma_{1} Re^{\gamma_{2}}), 
\end{align}
where $\gamma_1$, $\gamma_2$ are the nonlinear coefficients and $Re$ represents the instantaneous Reynolds number formulated as follows:
\begin{align}
  Re &= \frac{ \| {\bf v}_{\mathrm{w}}\| c_{\mathrm{m}}}{\nu}, \\
  Re_x &= \frac{v_{\mathrm{w},xz} c_{\mathrm{m}}}{\nu}, \\
  Re_y &= \frac{v_{\mathrm{w},yz} c_{\mathrm{m}}}{\nu}, 
\end{align}
where $c_{\mathrm{m}}$ is the mean chord length, and $\nu$ is the kinetic viscosity of air. By combining these, the complete expression of the conventional QSM can be written as follows: 
%
\begin{widetext}
  \begin{align}
  F_z &= F_{\mathrm{conv}}({\bf X}, \bm{\beta}, \bm{\gamma}) \notag \\
  &= \beta_{0} + \left(
    f_{\mathrm{tc}            } + 
    f_{\mathrm{tc}\mathrm{,Wg}}
  \right) \exp(\gamma_{\mathrm{tc1}} Re_x^{\gamma_{\mathrm{tc2}}}) \notag \\
  &\quad + \left(
    f_{\mathrm{td}            } + 
    f_{\mathrm{td}\mathrm{,Wg}}
  \right) \exp (\gamma_{\mathrm{td1}} Re_x^{\gamma_{\mathrm{td2}}}) \notag \\
  %
  %
  &\quad +
    f_{\mathrm{rc}            }
    \exp(\gamma_{\mathrm{rc1}} Re_x^{\gamma_{\mathrm{rc2}}}) \notag \\
  &\quad + 
    f_{\mathrm{rd}            }
    \exp(\gamma_{\mathrm{rd1}} Re_x^{\gamma_{\mathrm{rd2}}}) \notag \\
  &\quad + 
    f_{\mathrm{am}}
    (1 + \gamma_{\mathrm{am1}} Re^{\gamma_{\mathrm{am2}}}), \label{eq:conventional_QSM}
\end{align}

\begin{align*}
  f_{\mathrm{tc}            }
  &=
  \frac{1}{2} \rho 
  v_{\mathrm{w},xz}^2 S_{\mathrm{w}}
  C_L (\alpha_x, \beta_{\mathrm{tc1}}, \beta_{\mathrm{tc2}})
  \cos(\alpha_x),  \\
  f_{\mathrm{tc}\mathrm{,Wg}}
  &=
  \frac{1}{2} \rho   
  \left(
    v_{\mathrm{w},xz}(t_{\mathrm{eq}})
    C_L (\alpha_x(t_{\mathrm{eq}}), \beta_{\mathrm{tc,Wg1}}, \beta_{\mathrm{tc,Wg2}}) \right. \notag \\
    &\quad -
    \left.
    v_{\mathrm{w},xz}
    C_L (\alpha_x, \beta_{\mathrm{tc,Wg1}}, \beta_{\mathrm{tc,Wg2}})
  \right)
  \mathrm{sign}(a_{\mathrm{w},x}) \sqrt{c_{\mathrm{m}} |a_{\mathrm{w},x}|} \cos(\alpha_x) S_{\mathrm{w}}, \\
  f_{\mathrm{td}            }
  &=
  \frac{1}{2} \rho 
  v_{\mathrm{w},xz}^2 S_{\mathrm{w}}
  \left(
   \beta_{\mathrm{td}} \sin^2(\alpha_x) 
  \right) \sin(\alpha_x),  \\
  f_{\mathrm{td,Wg}}
  &=
  \frac{1}{2} \rho 
  \beta_{\mathrm{td,Wg}} \left(
    v_{\mathrm{w},xz}(t_{\mathrm{eq}}) \sin^2(\alpha_x(t_{\mathrm{eq}}))  -
    v_{\mathrm{w},xz} \sin^2(\alpha_x)
  \right)
  \mathrm{sign}(a_{\mathrm{w},x}) \sqrt{c_{\mathrm{m}} |a_{\mathrm{w},x}|} \sin(\alpha_x) S_{\mathrm{w}}, \\
  f_{\mathrm{rc}            }
  &=
  - \beta_{\mathrm{rc}} \rho v_{\mathrm{w},x} \omega_{\mathrm{w},y} S_{c}, \\
  f_{\mathrm{rd}}
  &= \frac{1}{2} \rho
  \beta_{\mathrm{rd}} \omega_{\mathrm{w},y} | \omega_{\mathrm{w},y} | S_{x|x|},\\
  f_{\mathrm{am}}
  &=
  \frac{\pi}{4} \rho  
  (
  - \beta_{\mathrm{am,trans}} S_{c} a_{\mathrm{w},z} 
  + \beta_{\mathrm{am,rot}} S_{x_{m}c} \psi_{\mathrm{w},y}
  ),
\end{align*}

  \begin{equation}
    C_L (\alpha, \beta_1, \beta_2) = 
    \beta_1 \alpha \left( 
      \alpha - \frac{\pi}{2}
    \right) +
    \beta_2 \alpha^2 \left( 
      \alpha - \frac{\pi}{2}
    \right), \notag 
  \end{equation}
\end{widetext}
%
where ${\bf X}$ is the kinematic parameter, and $\bm{\beta}$ and $\bm{\gamma}$ are the linear and nonlinear coefficients, respectively. The coefficients of the model were tuned by fitting with the correlation between the wing kinematics and aerodynamic forces derived by CFD using the Levenberg-Marquardt algorithm. To avoid overfitting, the cost function for the algorithm is defined as
\begin{equation}
  \frac{1}{N} 
  \left\| 
    {\bf F}_{\mathrm{conv}}({\bf X}, \bm{\beta}, \bm{\gamma}) - {\bf y} 
  \right\|_2^2
  + \lambda (\| \bm{\beta} \|_2^2 + \| \bm{\gamma} \|_2^2), 
\end{equation}
where the first term is an error of mean square, and the second term is a regularization term with a relative weight $\lambda$. ${\bf y}$ and ${\bf F}_{\mathrm{conv}}({\bf X}, \bm{\beta}, \bm{\gamma}) \in \mathbb{R}^N$ are vectors composed of $N$ aerodynamic force data computed by CFD and QSM, respectively.

In Sec. \ref{sec:1_introduction}, aerodynamic force estimation and fitting were performed for the hawkmoth forward flight \cite{Willmott1997-ow} ($v_{\mathrm{body}} = 2.1 \mathrm{ms}^{-1}$) using the conventional QSM (Figs. \ref{fig:10_conventional_QSM}(c) and (d)). In the estimation, the model's coefficients were determined by the training data described in Appendix \ref{sec:A3_App_dataDrivenFormulation}. For the fitting, the coefficients were tuned by the measured kinematics and associated CFD forces of the hawkmoth forward flight ($v_{\mathrm{body}} = 0 \mathrm{ms}^{-1}$ to $5 \mathrm{ms}^{-1}$), one of which was identical to the test data (shown in Fig. \ref{fig:10_conventional_QSM}(b)). In both cases, the conventional QSM deviates from the numerical data for some duration, indicating that other important aerodynamic mechanisms are not considered in the conventional QSM.

%
\section{\label{sec:A3_App_dataDrivenFormulation}DATA-DRIVEN FORMULATION}

%
\subsection{\label{subsec:App_wingKinDataset}Wing Kinematics Dataset}

\begin{table}[tbp]
  \caption{\label{tab:3_std1029_param}%
  Wing and body kinematics for the construction of training data. HM and FF represent the hawkmoth and fruit fly, respectively.
  }
  \begin{ruledtabular}
    \begin{tabular}{cc}
      $f$ (Hz) & 25 (HM) \\
       & 188 (FF) \\
      $A_{\mathrm{pos}}$ (deg) & 40--80 in step of 20  \\
      $A_{\mathrm{ele}}$ (deg)& 5--20 in step of 5 \\
      $A_{\mathrm{fea}}$ (deg)& 50--90 in step of 10 \\
      $\phi_{\mathrm{pos}}$ & 0.01, 0.3 ,0.6 ,0.9, 0.93, 0.96, 0.99 \\
      $\phi_{\mathrm{ele}}$ & $-1, 0, 1$ \\
      $\phi_{\mathrm{fea}}$ & 0.60--0.95 in step of 0.05 \\
      $\delta_{\mathrm{fea}}$ (rad)& $-\pi/9$--$\pi/9$ in step of $\pi/18$ \\
      $v_{\mathrm{fwd,m}} \mathrm{(ms^{-1})}$ & 0--6 in step of 2 (HM) \\
       & 0--0.9 in step of 0.3 (FF) \\
      $v_{\mathrm{lat,m}} \mathrm{(ms^{-1})}$ & 0 \\
      $\theta_{\mathrm{SP,m}}$ (deg) & Eq. \ref{eq:SPAestimate} \\
      $A_{\mathrm{fwd}} \mathrm{(ms^{-1})}$ & 1 (HM) \\
       & 0.15 (FF) \\
      $A_{\mathrm{lat}} \mathrm{(ms^{-1})}$ & $v_{\mathrm{fwd,m}}/2$\\
      $A_{\mathrm{SP}}$ (deg) & 25 \\
      $f_{\mathrm{fwd}}$ (Hz)& $f/9$ \\
      $f_{\mathrm{lat}}$ (Hz)& $f/13$ \\
      $f_{\mathrm{SP}}$ (Hz)& $f/16$ \\
      $\delta_{\mathrm{b}}$ (rad)& $\pi/3$--$6\pi/3$ in step of $\pi/3$ \\
    \end{tabular}
  \end{ruledtabular}
\end{table}

Three wing angles, namely positional $\theta_{\mathrm{pos}}$, elevation $\theta_{\mathrm{ele}}$, and feathering angle $\theta_{\mathrm{fea}}$, for the training were synthesized based on the measured flight data of various insects and formulation following \cite{Berman2007-tg}:
\begin{align}
  \theta_{\mathrm{pos}} &=
  \frac{A_{\mathrm{pos}}}{\arcsin \phi_{\mathrm{pos}}}
  \arcsin\left(
    \phi_{\mathrm{pos}} \sin\left(
      2\pi f t + \frac{\pi}{2}
    \right)
  \right), \notag \\
  &  \\
  \theta_{\mathrm{ele}} &= 
  \begin{cases}
    A_{\mathrm{ele}} \left(
      \cos (4 \pi f t) - 1
    \right) & \text{if $\phi_{\mathrm{ele}} = -1$} \\
    A_{\mathrm{ele}} \sin(2 \pi f t) & \text{if $\phi_{\mathrm{ele}} = 0$} \\
    A_{\mathrm{ele}} \sin(4 \pi f t) & \text{if $\phi_{\mathrm{ele}} = 1$},
  \end{cases}  \\
  \theta_{\mathrm{fea}} &= A_{\mathrm{fea}} K(\phi_{\mathrm{fea}}) \left(
    \phi_{\mathrm{fea}} \sin(2 \pi f t + \delta_{\mathrm{fea}}) 
  \right. \notag \\
    &\quad + \left.
    (1 - \phi_{\mathrm{fea}}) \sin \left(
      3(2 \pi f t + \delta_{\mathrm{fea}})
    \right)
  \right),  \\
  K(\phi_{\mathrm{fea}}) &=
  \begin{cases}
    - \sqrt{
      \frac{27(1-\phi_{\mathrm{fea}})}{(3-2\phi_{\mathrm{fea}})^3}
    } 
    & \text{if $\phi_{\mathrm{fea}} \leq 0.9$} \\
    - \frac{1}{2\phi_{\mathrm{fea}}-1} & \text{if $0.9 \leq \phi_{\mathrm{fea}} < 1.0$},
  \end{cases}
\end{align}

\begin{figure}[tbp]
    \centering
    \includegraphics[width=0.48\textwidth]{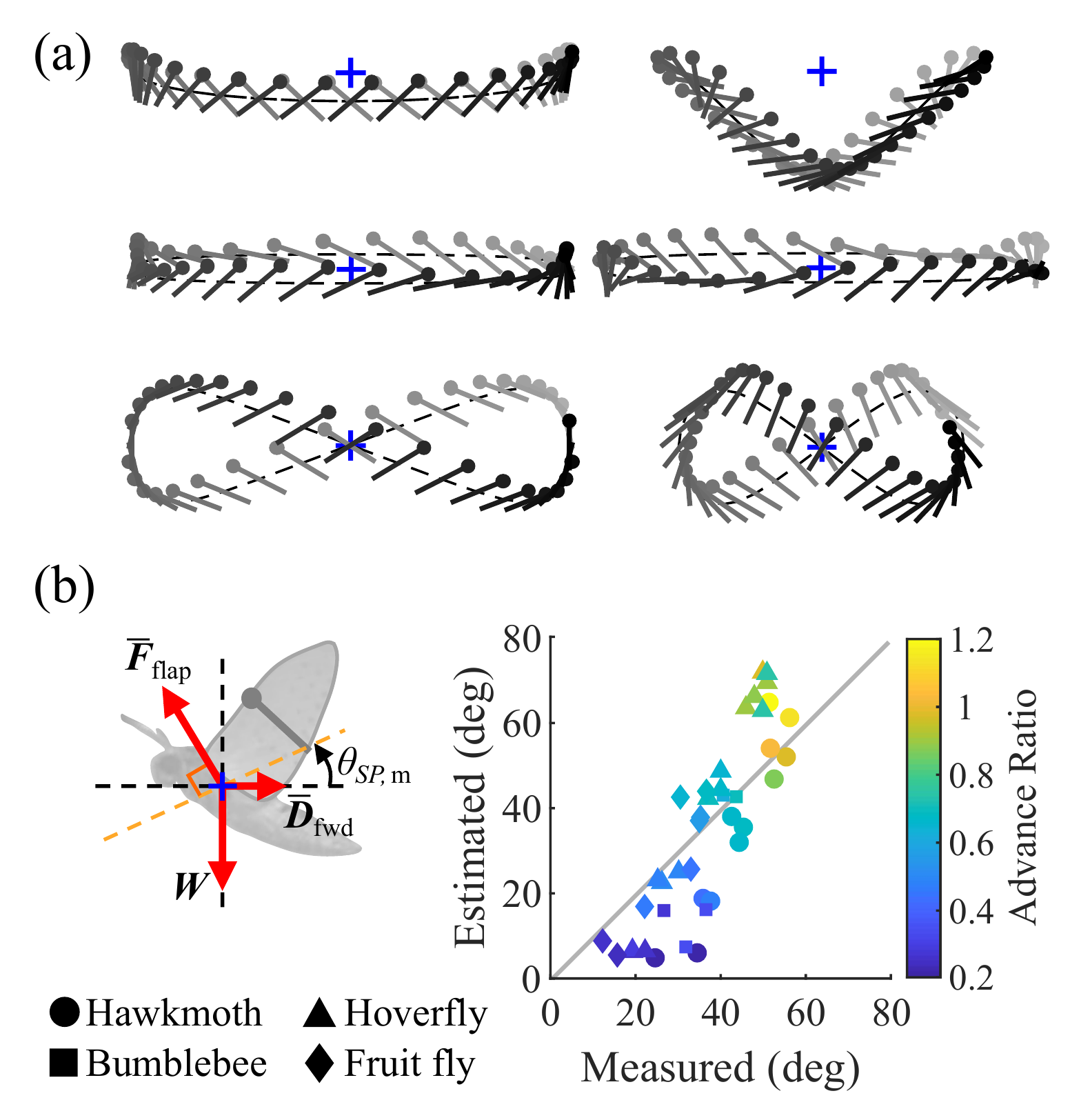}
    \caption{
      Generation of kinematic data for sparse identification and QSM construction. (a) Side view of an example of the wing kinematics. The lines and circles represent the wing cross-sections and leading edges, respectively. The blue crosses indicate the wing rotation axis. (b) SPA estimated using Eq. C9 based on the assumption that the force $\bar{{\bf F}}_{\mathrm{flap}}$ generated perpendicular to the stroke plane balances the drag $\bar{{\bf D}}_{\mathrm{fwd}}$ and body weight ${\bf W}$ (left). The values align with the measured values. The circles, triangles, squares, and diamonds represent data from a hawkmoth \cite{Willmott1997-ow}, hoverfly \cite{Meng2016-kn}, bumblebee \cite{Dudley1990-ve}, and fruit fly \cite{Zhu2020-za}, respectively.
    }
    \label{fig:100_training_data_kin}
\end{figure}

where $f$ is the wingbeat frequency, $A_{\mathrm{pos}}$, $A_{\mathrm{ele}}$, $A_{\mathrm{fea}}$ are the wingbeat amplitudes, $\delta_{\mathrm{fea}}$ is the phase difference of the feathering angle relative to the positional angle, and $\phi_{\mathrm{pos}}$, $\phi_{\mathrm{ele}}$, $\phi_{\mathrm{fea}}$ are the parameters used to determine the waveforms (Fig. \ref{fig:20_wingkin_morphology}(f)). $f$ was set to 25 Hz and 188 Hz for the hawkmoth and fruit flies, respectively. The other parameters were systematically varied, as listed in Table \ref{tab:3_std1029_param}. Because one set of parameters corresponds to one wingbeat, the number of wingbeats in the training data becomes substantial, reaching as many as 56700 wingbeats if all the combinations of the parameters in the table are used. However, because many of the kinematics can be interpolated by others, we selected only 50 wingbeats using sequential forward selection \cite{Whitney1971-hj} that sequentially chooses a subset based on an evaluation function. The evaluation function was defined by partitioning the 6-dimensional space composed of angular velocities and accelerations of the wings into discrete cells, and counting how many of these occupied cells were covered by the selected wingbeats. Therefore, the selected wingbeats were chosen to cover the space of angular velocities and accelerations as uniformly as possible, resulting in diverse trajectories (Fig. \ref{fig:100_training_data_kin}(a)). 

\begin{figure*}[tbp]
    \centering
    \includegraphics[width=1.0\textwidth]{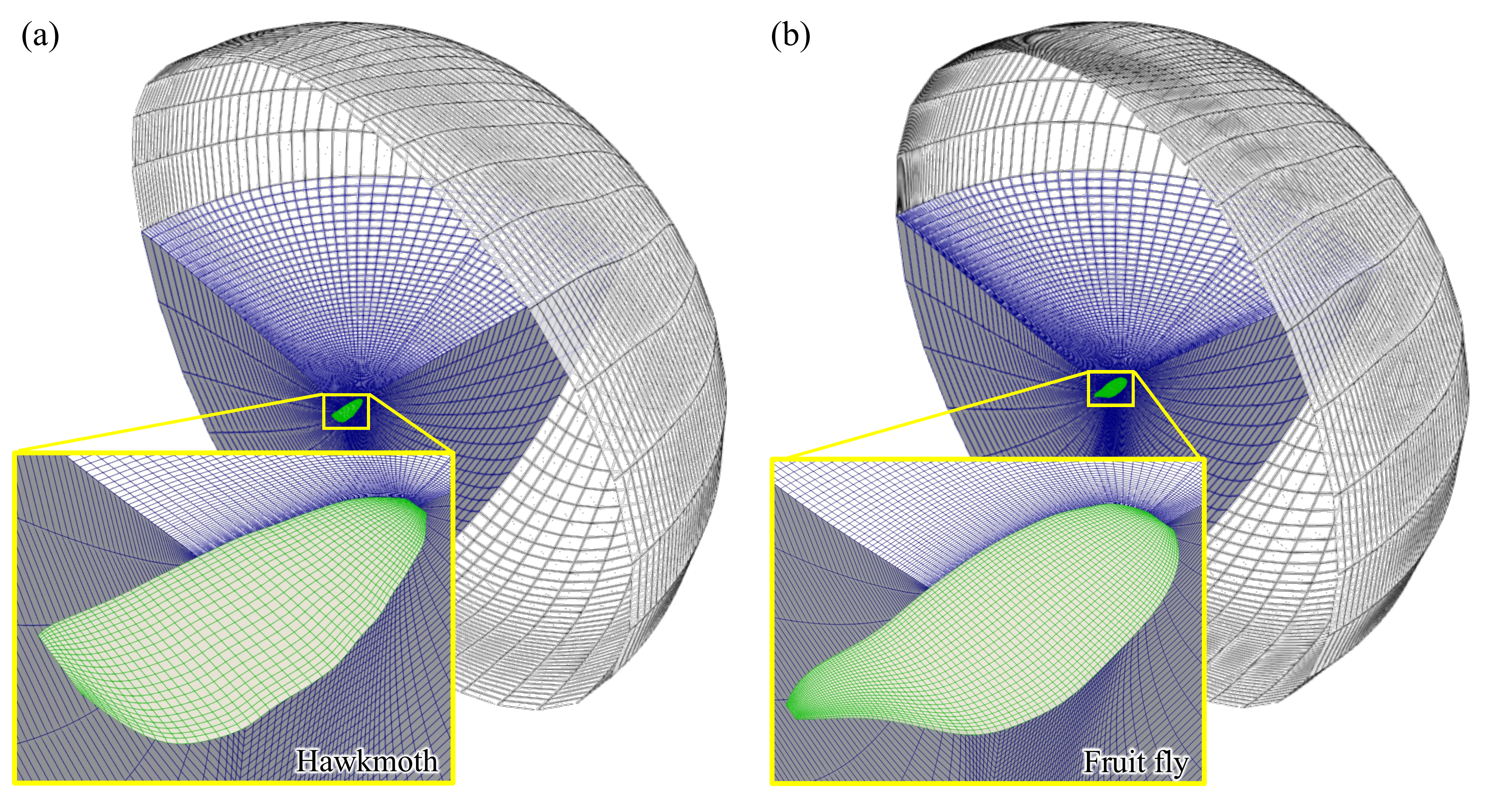}
    \caption{
      Grids around the wing models of the (a) hawkmoth and (b) fruit fly.
    }
    \label{fig:110_wing_grids}
\end{figure*}

The forward velocity $v_{\mathrm{fwd}}$, lateral velocity $v_{\mathrm{lat}}$, and stroke plane angle (SPA) $\theta_{\mathrm{SP}}$ relative to the horizontal plane were synthesized as
\begin{align}
  v_{\mathrm{fwd}}(t) &= v_{\mathrm{fwd,m}}
  + A_{\mathrm{fwd}} \sin(2 \pi f_{\mathrm{fwd}} t + \delta_{\mathrm{b}}), \\
  v_{\mathrm{lat}}(t) &= v_{\mathrm{lat,m}}
  + A_{\mathrm{lat}} \sin(2 \pi f_{\mathrm{lat}} t + \delta_{\mathrm{b}}), \\
  \theta_{\mathrm{SP}}(t) &= \theta_{\mathrm{SP,m}}
  + A_{\mathrm{SP}} \sin(2 \pi f_{\mathrm{SP}} t + \delta_{\mathrm{b}}),
\end{align}
where $v_{\mathrm{fwd,m}}$ is the mean forward velocity; $v_{\mathrm{lat,m}}$ is the mean lateral velocity; $\theta_{\mathrm{SP,m}}$ is the mean SPA; $A_{\mathrm{fwd}}$, $A_{\mathrm{lat}}$, $A_{\mathrm{SP}}$ are the amplitudes; $f_{\mathrm{fwd}}$, $f_{\mathrm{lat}}$, $f_{\mathrm{SP}}$ are the frequencies; and $\delta_{\mathrm{b}}$ is the phase parameter of the waveforms (Fig. \ref{fig:20_wingkin_morphology}(g)). These parameters were varied as shown in Table \ref{tab:3_std1029_param}. $v_{\mathrm{fwd,m}}$, $A_{\mathrm{fwd}}$, and $A_{\mathrm{lat}}$ were determined so as to cover the flight speed of each insect. Different frequencies were applied to the forward and lateral velocities and SPAs to include their various combinations. The wing attitude variations caused by the roll, pitch, and yaw motion of the body were set to zero, assuming that they can be covered by the variations of $\theta_{\mathrm{pos}}$, $\theta_{\mathrm{ele}}$, and $\theta_{\mathrm{fea}}$.

Because flapping wings mainly generate an aerodynamic force normal to the stroke plane when averaged over a cycle, the SPA of insects, as in helicopters, increases as the forward velocity increases to generate thrust. To avoid the training data that deviate significantly from the insect wing kinematics, we investigated the kinematic correlation between the forward velocity and SPA and formulated the mean SPA $\theta_{\mathrm{SP,m}}$ for the training data based on the following condition: The cycle-averaged aerodynamic force normal to the wing surface generated by only flapping 
$\bar{{\bf F}}_{\mathrm{flap}}$ is balanced by the horizontal force $\bar{{\bf D}}_{\mathrm{fwd}}$
on the wing generated by the body motion and body weight ${\bf W}$ (Fig. \ref{fig:100_training_data_kin}(b)). The magnitude of 
$\bar{{\bf D}}_{\mathrm{fwd}}$
was approximated by the drag acting on the wing translating with velocity $v_{\mathrm{fwd,m}}$ and an angle of attack of $35^{\circ}$ as follows:
\begin{equation}
  |\bar{{\bf D}}_{\mathrm{fwd}}| = 0.5 \rho v_{\mathrm{fwd,m}}^2 S_{\mathrm{2w}} C_{D},
\end{equation}
where $\rho$ is air density, and $S_{\mathrm{2w}}$ is the wing area of two wings. This angle of attack was determined based on previous measurements \cite{Ellington1984-kv} and various measurements of insect kinematics \cite{Zhu2020-za,Meng2016-kn}. The drag coefficient $C_D$ was defined as 1.2 based on the experiments with revolving wing \cite{Dickinson1993-cp}. Therefore, the mean SPA can be calculated as follows:
\begin{equation}
  \theta_{\mathrm{SP,m}} = \frac{180}{\pi} 
  \arctan \left(
    \frac
    {0.5 \rho v_{\mathrm{fwd,m}}^2 S_{\mathrm{2w}} C_{\mathrm{D}}}
    {m_{\mathrm{b}} g}
  \right), \label{eq:SPAestimate}
\end{equation}
where $m_{\mathrm{b}}$ (kg) is the mass of the insect \cite{Muijres2014-hx,Willmott1997-yw}, and $g$ is the gravitational acceleration. As shown in Fig. \ref{fig:100_training_data_kin}(b), despite the rough approximation, the formulation captures the relationship between the forward velocity and SPA of the four kinds of insects \cite{Willmott1997-ow,Zhu2020-za,Meng2016-kn,Dudley1990-ve} with the Reynolds numbers ranging from $10^2$ to $10^4$. Therefore, we determined the stroke plane angle in the training data using Eq. \ref{eq:SPAestimate}.

Combining the wing motion relative to the stroke plane for 50 flaps, four forward velocities $v_{\mathrm{fwd,m}}$, and six phase parameters $\delta_{\mathrm{b}}$, we constructed a wing kinematics dataset consisting of data for 1,200 flaps and used it for training.

\begin{figure}[tbp]
    \centering
    \includegraphics[width=0.48\textwidth]{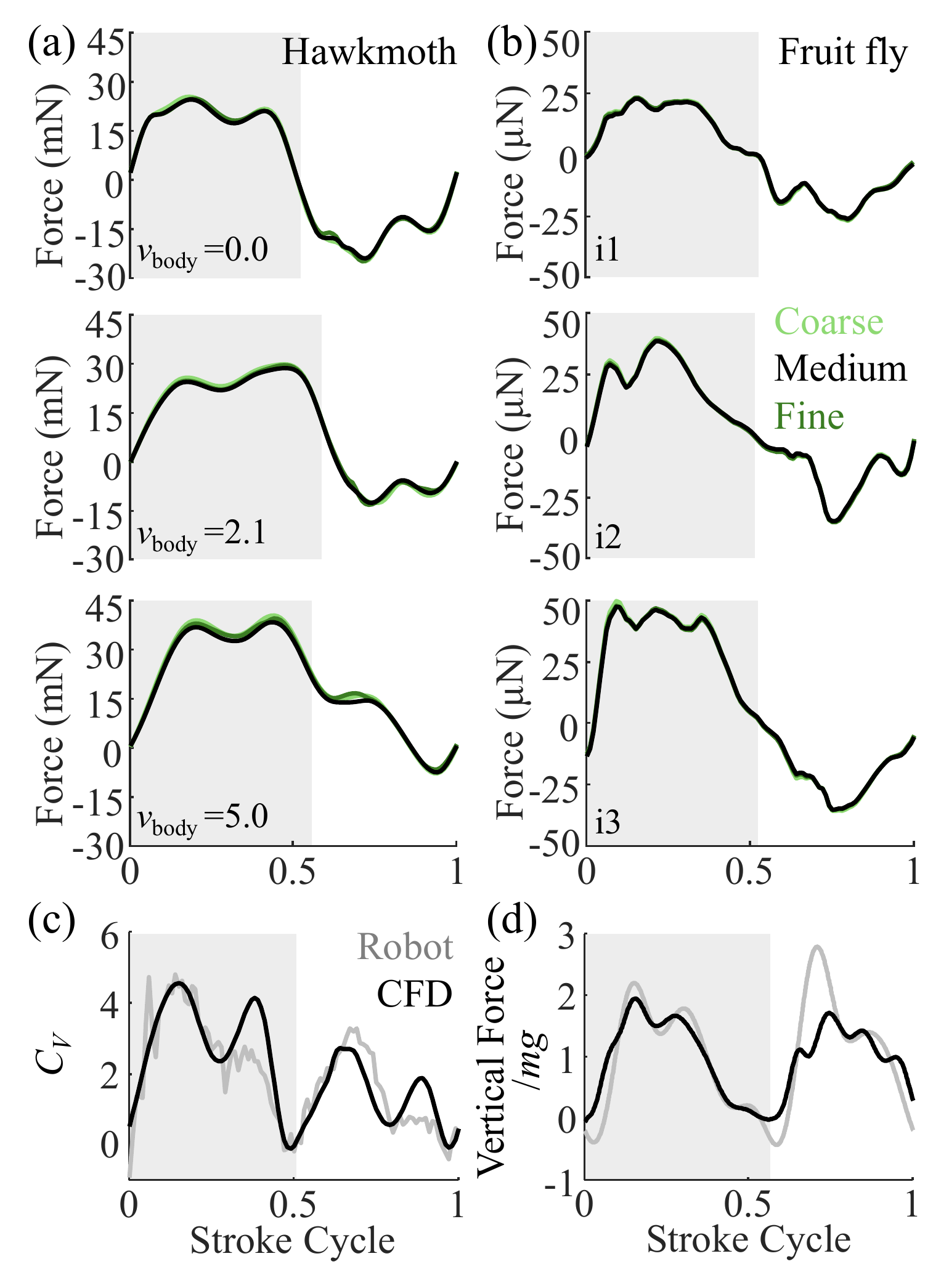}
    \caption{
      Validation and verification of the computational model for hawkmoth (a, c) and fruit fly (b, d) wings. (a, b) Grid independence verified by the closely matching results of the coarse, fine, and finer grids. Results obtained with the fine mesh are employed as the ground truth in this study. (c, d) Comparison between aerodynamic forces predicted by CFD and those measured in robot experiments \cite{Muijres2014-hx,Lua2010-vj}. The gray shaded areas represent the downstroke.
    }
    \label{fig:120_validation_CFD}
\end{figure}

%
\subsection{\label{subsec:App_validationCFD}Validation and Verification of the Computational Model}
The wing morphologies and computational domain were discretized using structured O-O grids with $45 \times 65 \times 101$ nodes for the hawkmoth and $55 \times 110 \times 101$ nodes for the fruit fly in the chordwise, spanwise, and outer directions, respectively (Fig. \ref{fig:110_wing_grids}). The distance between the wing surface and outermost grid points exceeds 25 $c_{\mathrm{m}}$ ($c_{\mathrm{m}}$ is the mean chord length). To investigate the grid dependence of aerodynamic force estimation, three cases were tested for both hawkmoths and fruit flies: coarse grids ($33\times 53\times 101$ for hawkmoths and $41\times 82\times 101$ for fruit flies), fine grids ($45\times 65\times 101$ for hawkmoths and $55\times 110\times 101$ for fruit flies), and finer grids ($55\times 89\times 101$ for hawkmoths and $67\times 134\times 101$ for fruit flies). Fig. \ref{fig:120_validation_CFD} shows that the computed aerodynamic forces show marginal discrepancies among the three cases (within 5\% between fine and finer grids for both wings), and the grids used in this study (fine grids) were sufficiently small to compute the self-consistent aerodynamic forces. 

The aerodynamic forces obtained from CFD analyses using these models show good agreement with those measured in robots.
The vertical forces by hovering hawkmoths \cite{Lua2010-vj} and maneuvering fruit flies \cite{Muijres2014-hx} match well during the downstroke (Fig. \ref{fig:120_validation_CFD}(c, d)).
Although a discrepancy is observed in the peak values during the upstroke, the errors in cycle-averaged forces remain low for both cases (hawkmoth: 7.4 \%, fruit fly: 2.5 \%), indicating that the impact is minor. Therefore, the CFD results in this study are considered sufficiently accurate to serve as ground truth.

%
\subsection{\label{subsec:App_candidateFuncs}Candidate Functions}
The candidate function library was constructed from the first, second, and third polynomials of ``element functions''. These functions are composed of the wing's angles of attack $\alpha_x$ and $\alpha_y$, velocity ${\bf v}_{\mathrm{w}}$, acceleration ${\bf a}_{\mathrm{w}}$, angular velocity $\bm{\omega}_{\mathrm{w}}$, angular acceleration $\bm{\psi}_{\mathrm{w}}$, and body velocity ${\bf v}_{\mathrm{b}}$. Specifically, in the case of angle of attack $\alpha_x$ (and $\alpha_y$), the element function contains 
$\sin(\alpha_x)$, $\cos(\alpha_x)$, $\sin(2\alpha_x)$, $\cos(2\alpha_x)$, $\sin^3(\alpha_x)$, and $\cos^3(\alpha_x)$.
For wing velocity ${\bf v}_{\mathrm{w}}$, they contain the $x$, $y$, and $z$ components in the wing-fixed frame 
$v_{\mathrm{w},x}$, $v_{\mathrm{w},y}$, and $v_{\mathrm{w},z}$; their absolute values $|v_{\mathrm{w},x}|$, $|v_{\mathrm{w},y}|$, and $|v_{\mathrm{w},z}|$, and the magnitude of the vector $|{\bf v}_{\mathrm{w}}|$ (same for ${\bf a}_{\mathrm{w}}$, $\bm{\omega}_{\mathrm{w}}$, $\bm{\psi}_{\mathrm{w}}$, and ${\bf v}_{\mathrm{b}}$). For wing acceleration, the signed square roots 
$\mathrm{sign}(a_{\mathrm{w},x})\sqrt{|a_{\mathrm{w},x}|}$, $\mathrm{sign}(a_{\mathrm{w},y})\sqrt{|a_{\mathrm{w},y}|}$, and $\mathrm{sign}(a_{\mathrm{w},z})\sqrt{|a_{\mathrm{w},z}|}$
were included to address the Wagner effect and its related effects. Redundant functions that were linearly dependent or had equivalent effects were removed to ensure linear independence among the candidates. The time dimension was restricted to ($\mathrm{s}^0$), ($\mathrm{s}^{-1}$), or ($\mathrm{s}^{-2}$) because the other dimensions were considered physically unrealistic. The candidate function library consisted of 5,548 functions.

%
\subsection{\label{subsec:App_featureelim}Feature Elimination}
Because manual determination of the new aerodynamic mechanisms from thousands of candidate functions is not feasible, we repeated an operation similar to STRidge \cite{Rudy2017-fh} to identify the important functions, as described in Sec. \ref{subsubsec:modelDiscovery}. In this study, RIDGE regression minimizes
\begin{align}
  \frac{1}{N} 
  \left\| 
    \bm{\Theta}({\bf X})\bm{\beta} - {\bf y} 
  \right\|_2^2
  + \lambda \| \bm{\beta} \|_2^2, 
\end{align}
where ${\bf y}$ is the aerodynamic force computed by CFD, $\bm{\Theta}({\bf X})$ is a matrix with standardized candidate functions in each column as in \cite{Brunton2016-ld}, $\bm{\beta}$ is the regression coefficient, and $\lambda$ is the hyper parameter for determining the weight of the regularization term (the second term). The terms in conventional QSM (Eq. \ref{eq:conventional_QSM}) were also included in $\bm{\Theta}({\bf X})$ to investigate the aerodynamic mechanisms not considered in conventional QSM, and all the linear coefficients were tuned simultaneously. We randomly selected half of the training data as validation data and selected the optimal hyperparameter from 10 values ranging from $10^{-15}$ to $10^1$ based on its fitness to the validation data.

To calculate the importance index of each candidate function, we referred to the magnitude of each element in $\bm{\beta}$, which represents the relative importance of the candidate functions in $\bm{\Theta}({\bf X})$. In STLS \cite{Brunton2016-ld} and STRidge \cite{Rudy2017-fh}, the functions to be eliminated are selected using only one regression. In our study, to deal with the limitation of data size owing to numerous candidate functions, we repeated random sampling and regression $N_{\mathrm{ave}}$ times and computed the importance index $I_j$ of the $j$th candidate function using
\begin{align}
  I_j = \sum_{\ell=1}^{N_{\mathrm{ave}}}
  \frac{\beta_{\ell, j}}{\| \bm{\beta}_{\ell} \|_1}, 
\end{align}
where $\bm{\beta}_l ~ (l = 1, \ldots, N_{\mathrm{ave}})$ is the regression coefficient of the $l$th regression, and $\bm{\beta}_{l, j}$ is the $j$th element of $\bm{\beta}_l$. $N_{\mathrm{elim}}$ functions with the lowest indices were eliminated. In this work, we set $N_{\mathrm{ave}} = 10$ and varied $N_{\mathrm{elim}}$ according to the number of the remaining candidate functions $N_{\mathrm{pred}}$, where $N_{\mathrm{elim}} = 0.02 N_{\mathrm{pred}}$ when $1000 < N_{\mathrm{pred}}$ and $N_{\mathrm{elim}} = \mathrm{max}(1, 0.01 N_{\mathrm{pred}})$ when $N_{\mathrm{pred}} \leq 1000$.

%
\section{\label{sec:A4_App_qsmWithNovel}QSM WITH NOVEL MECHANISMS}

%
\subsection{\label{subsec:App_dataDrivenQsm}Data-Driven QSM}
Incorporating the three identified mechanisms, the new QSM is written as follows:
%
\begin{widetext}
  \begin{align}
  F_z &= F_{\mathrm{new}}({\bf X}, \bm{\beta}, \bm{\gamma}) \notag \\
  &= \beta_{0} + 
  \left(
    f_{\mathrm{tc,ch,bd}} + 
    f_{\mathrm{tc,ch,cp}} + 
    f_{\mathrm{tc,ch,fl}} + 
    f_{\mathrm{tc,ch,Wg}}
  \right) \exp(\gamma_{\mathrm{tc,ch1}} Re_x^{\gamma_{\mathrm{tc,ch2}}}) \notag \\
  &\quad + \left(
    f_{\mathrm{td,ch,bd}} + 
    f_{\mathrm{td,ch,cp}} + 
    f_{\mathrm{td,ch,fl}} +
    f_{\mathrm{td,ch,Wg}}
  \right) \exp (\gamma_{\mathrm{td,ch1}} Re_x^{\gamma_{\mathrm{td,ch2}}}) \notag \\
  &\quad + \left(
    f_{\mathrm{rc,ch,bd}} + 
    f_{\mathrm{rc,ch,fl}} +
    f_{\mathrm{rc,ch,Wg}}
  \right) \exp(\gamma_{\mathrm{rc,ch1}} Re_x^{\gamma_{\mathrm{rc,ch2}}}) \notag \\
  &\quad + 
    f_{\mathrm{rd,ch}} 
    \exp(\gamma_{\mathrm{rd,ch1}} Re_x^{\gamma_{\mathrm{rd,ch2}}}) \notag \\
  &\quad + 
    f_{\mathrm{am}} 
    (1 + \gamma_{\mathrm{am1}} Re^{\gamma_{\mathrm{am2}}}) \notag \\
  &\quad + 
  \left(
    f_{\mathrm{tc,sp,bd}} + 
    f_{\mathrm{tc,sp,cp}} + 
    f_{\mathrm{tc,sp,fl}} + 
    f_{\mathrm{tc,sp,Wg}}
  \right) \exp(\gamma_{\mathrm{tc,sp1}} Re_y^{\gamma_{\mathrm{tc,sp2}}}) \notag \\
  &\quad + \left(
    f_{\mathrm{td,sp,bd}} + 
    f_{\mathrm{td,sp,cp}} + 
    f_{\mathrm{td,sp,fl}} +
    f_{\mathrm{td,sp,Wg}}
  \right) \exp (\gamma_{\mathrm{td,sp1}} Re_y^{\gamma_{\mathrm{td,sp2}}}) \notag \\
  &\quad + \left(
    f_{\mathrm{rc,sp,bd}} + 
    f_{\mathrm{rc,sp,fl}} +
    f_{\mathrm{rc,sp,Wg}}
  \right) \exp(\gamma_{\mathrm{rc,sp1}} Re_y^{\gamma_{\mathrm{rc,sp2}}}) \notag \\
  &\quad + 
    f_{\mathrm{rd,sp}} ~
    \exp(\gamma_{\mathrm{rd,sp1}} Re_y^{\gamma_{\mathrm{rd,sp2}}}) \label{eq:new_QSM}
  %
  %
\end{align}

\begin{align*}
  f_{\mathrm{tc,ch,bd}}
  &=
  \frac{1}{2} \rho 
  v_{\mathrm{b},xz}^2 S_{\mathrm{w}}
  C_L (\alpha, \beta_{\mathrm{tc,ch,bd1}}, \beta_{\mathrm{tc,ch,bd2}})
  \cos(\alpha_x)  \\
  f_{\mathrm{tc,ch,fl}}
  &=
  \frac{1}{2} \rho 
  \left(
    \omega_{\mathrm{w},xz}^2 S_{yy} + 
    \omega_{\mathrm{w},y}^2 S_{xx} -
    2 \omega_{\mathrm{w},x} \omega_{\mathrm{w},y} S_{xy}
  \right) 
  C_L (\alpha, \beta_{\mathrm{tc,ch,fl1}}, \beta_{\mathrm{tc,ch,fl2}})
  \cos(\alpha_x)  \\
  f_{\mathrm{tc,ch,cp}}
  &=
  \frac{1}{2} \rho 
  \left(
    2 (v_{\mathrm{b},z} \omega_{\mathrm{w},x} - v_{\mathrm{b},x} \omega_{\mathrm{w},z}) S_{y} -
    2 v_{\mathrm{b},z} \omega_{\mathrm{w},y} S_{x}
  \right) 
  C_L (\alpha, \beta_{\mathrm{tc,ch,cp1}}, \beta_{\mathrm{tc,ch,cp2}})
  \cos(\alpha_x)  \\
  f_{\mathrm{tc,ch,Wg}}
  &=
  \frac{1}{2} \rho 
  \left(
    v_{\mathrm{w},xz}(t_{\mathrm{eq}})
    C_L (\alpha_x(t_{\mathrm{eq}}), \beta_{\mathrm{tc,ch,Wg1}}, \beta_{\mathrm{tc,ch,Wg2}}) \right. \notag \\
    &\quad -
    \left.
    v_{\mathrm{w},xz}
    C_L (\alpha_x, \beta_{\mathrm{tc,ch,Wg1}}, \beta_{\mathrm{tc,ch,Wg2}})
  \right)
  \mathrm{sign}(a_{\mathrm{w},x}) \sqrt{c_{\mathrm{m}} |a_{\mathrm{w},x}|} \cos(\alpha_x) S_{\mathrm{w}} \\
  f_{\mathrm{td,ch,bd}}
  &=
  \frac{1}{2} \rho 
  v_{\mathrm{b},xz}^2 S_{\mathrm{w}}
  \left(
   \beta_{\mathrm{td,ch,bd}} \sin^2(\alpha_x) 
  \right) \sin(\alpha_x)  \\
  f_{\mathrm{td,ch,fl}}
  &=
  \frac{1}{2} \rho 
  \left(
    \omega_{\mathrm{w},xz}^2 S_{yy} + 
    \omega_{\mathrm{w},y}^2 S_{xx} -
    2 \omega_{\mathrm{w},x} \omega_{\mathrm{w},y} S_{xy}
  \right) 
  \left(
    \beta_{\mathrm{td,ch,fl}} \sin^2(\alpha_x) 
  \right) \sin(\alpha_x)  \\
  f_{\mathrm{td,ch,cp}}
  &=
  \frac{1}{2} \rho 
  \left(
    2 (v_{\mathrm{b},z} \omega_{\mathrm{w},x} - v_{\mathrm{b},x} \omega_{\mathrm{w},z}) S_{y} -
    2 v_{\mathrm{b},z} \omega_{\mathrm{w},y} S_{x}
  \right) 
  \left(
    \beta_{\mathrm{td,ch,cp}} \sin^2(\alpha_x) 
  \right) \sin(\alpha_x)  \\
  f_{\mathrm{td,ch,Wg}}
  &=
  \frac{1}{2} \rho 
  \beta_{\mathrm{td,ch,Wg}} \left(
    v_{\mathrm{w},xz}(t_{\mathrm{eq}}) \sin^2(\alpha_x(t_{\mathrm{eq}}))  -
    v_{\mathrm{w},xz} \sin^2(\alpha_x)
  \right)
  \mathrm{sign}(a_{\mathrm{w},x}) \sqrt{c_{\mathrm{m}} |a_{\mathrm{w},x}|} \sin(\alpha_x) S_{\mathrm{w}} \\
  f_{\mathrm{rc,ch,bd}}
  &= 
  -
  \beta_{\mathrm{rc,ch,bd}} \rho v_{\mathrm{b},x} \omega_{\mathrm{w},y} S_{c} \\
  f_{\mathrm{rc,ch,fl}}
  &=
  \beta_{\mathrm{rc,ch,fl}} \rho \omega_{\mathrm{w},z} \omega_{\mathrm{w},y} S_{cy} \\
  f_{\mathrm{rc,ch,Wg}}
  &=
  \beta_{\mathrm{rc,ch,Wg}} \rho \left(
    \omega_{\mathrm{w},y}(t_{\mathrm{eq}}) - \omega_{\mathrm{w},y}
  \right) \mathrm{sign}(a_{\mathrm{w},x}) \sqrt{c_{\mathrm{m}} |a_{\mathrm{w},x}|} S_{c} \\
  f_{\mathrm{rd,ch}}
  &= \frac{1}{2} \rho
  \beta_{\mathrm{rd,ch}} \omega_{\mathrm{w},y} | \omega_{\mathrm{w},y} | S_{x|x|}\\
  f_{\mathrm{am}}
  &=
  \frac{\pi}{4} \rho  
  (
  - \beta_{\mathrm{am,trans}} S_{c} a_{\mathrm{w},z} 
  + \beta_{\mathrm{am,rot}} S_{x_{m}c} \psi_{\mathrm{w},y}
  ) \\
  f_{\mathrm{tc,sp,bd}}
  &=
  \frac{1}{2} \rho 
  v_{\mathrm{b},yz}^2 S_{\mathrm{w}}
  C_L (\alpha, \beta_{\mathrm{tc,sp,bd1}}, \beta_{\mathrm{tc,sp,bd2}})
  \cos(\alpha_y)  \\
  f_{\mathrm{tc,sp,fl}}
  &=
  \frac{1}{2} \rho 
  \left(
    \omega_{\mathrm{w},yz}^2 S_{xx} + 
    \omega_{\mathrm{w},x}^2 S_{yy} -
    2 \omega_{\mathrm{w},x} \omega_{\mathrm{w},y} S_{xy}
  \right) 
  C_L (\alpha, \beta_{\mathrm{tc,sp,fl1}}, \beta_{\mathrm{tc,sp,fl2}})
  \cos(\alpha_y)  \\
  f_{\mathrm{tc,sp,cp}}
  &=
  \frac{1}{2} \rho 
  \left(
    2 (v_{\mathrm{b},y} \omega_{\mathrm{w},z} - v_{\mathrm{b},z} \omega_{\mathrm{w},y}) S_{x} +
    2 v_{\mathrm{b},z} \omega_{\mathrm{w},x} S_{y}
  \right) 
  C_L (\alpha, \beta_{\mathrm{tc,sp,cp1}}, \beta_{\mathrm{tc,sp,cp2}})
  \cos(\alpha_y)  \\
  f_{\mathrm{tc,sp,Wg}}
  &=
  \frac{1}{2} \rho 
  \left(
    v_{\mathrm{w},yz}(t_{\mathrm{eq}})
    C_L (\alpha_y(t_{\mathrm{eq}}), \beta_{\mathrm{tc,sp,Wg1}}, \beta_{\mathrm{tc,sp,Wg2}}) \right. \notag \\
    &\quad -
    \left.
    v_{\mathrm{w},yz}
    C_L (\alpha_y, \beta_{\mathrm{tc,sp,Wg1}}, \beta_{\mathrm{tc,sp,Wg2}})
  \right)
  \mathrm{sign}(a_{\mathrm{w},y}) \sqrt{R |a_{\mathrm{w},y}|} \cos(\alpha_x) S_{\mathrm{w}} \\
  f_{\mathrm{td,sp,bd}}
  &=
  \frac{1}{2} \rho 
  v_{\mathrm{b},yz}^2 S_{\mathrm{w}}
  \left(
   \beta_{\mathrm{td,sp,bd}} \sin^2(\alpha_y) 
  \right) \sin(\alpha_y)  \\
  f_{\mathrm{td,sp,fl}}
  &=
  \frac{1}{2} \rho 
  \left(
    \omega_{\mathrm{w},yz}^2 S_{xx} + 
    \omega_{\mathrm{w},x}^2 S_{yy} -
    2 \omega_{\mathrm{w},x} \omega_{\mathrm{w},y} S_{xy}
  \right) 
  \left(
    \beta_{\mathrm{td,sp,fl}} \sin^2(\alpha_y) 
  \right) \sin(\alpha_y)  \\
  f_{\mathrm{td,sp,cp}}
  &=
  \frac{1}{2} \rho 
  \left(
    2 (v_{\mathrm{b},y} \omega_{\mathrm{w},z} - v_{\mathrm{b},z} \omega_{\mathrm{w},y}) S_{x} +
    2 v_{\mathrm{b},z} \omega_{\mathrm{w},x} S_{y}
  \right) 
  \left(
    \beta_{\mathrm{td,sp,cp}} \sin^2(\alpha_y) 
  \right) \sin(\alpha_y)  \\
  f_{\mathrm{td,sp,Wg}}
  &=
  \frac{1}{2} \rho 
  \beta_{\mathrm{td,sp,Wg}} \left(
    v_{\mathrm{w},yz}(t_{\mathrm{eq}}) \sin^2(\alpha_y(t_{\mathrm{eq}}))  -
    v_{\mathrm{w},yz} \sin^2(\alpha_y)
  \right)
  \mathrm{sign}(a_{\mathrm{w},y}) \sqrt{R |a_{\mathrm{w},y}|} \sin(\alpha_y) S_{\mathrm{w}} \\
  f_{\mathrm{rc,sp,bd}}
  &=
  \beta_{\mathrm{rc,sp,bd}} \rho v_{\mathrm{b},y} \omega_{\mathrm{w},x} S_{s} \\
  f_{\mathrm{rc,sp,fl}}
  &=
  \beta_{\mathrm{rc,sp,fl}} \rho \omega_{\mathrm{w},z} \omega_{\mathrm{w},x} S_{sx} \\
  f_{\mathrm{rc,sp,Wg}}
  &=
  \beta_{\mathrm{rc,sp,Wg}} \rho \left(
    \omega_{\mathrm{w},x}(t_{\mathrm{eq}}) - \omega_{\mathrm{w},x}
  \right) \mathrm{sign}(a_{\mathrm{w},y}) \sqrt{R |a_{\mathrm{w},y}|} S_{s} \\
  f_{\mathrm{rd,sp}}
  &= \frac{1}{2} \rho
  \beta_{\mathrm{rd,sp}} \omega_{\mathrm{w},x} | \omega_{\mathrm{w},x} | S_{y|y|}\\
\end{align*}

  \noindent where
  \begin{gather}
    S_x = 
    \iint_{\mathrm{wing}} x~ \mathrm{d}x \mathrm{d}y, ~~~
    S_y = 
    \iint_{\mathrm{wing}} y~ \mathrm{d}x \mathrm{d}y, ~~~
    S_{cy} =
    \iint_{\mathrm{wing}} c(y)y~ \mathrm{d}x \mathrm{d}y, ~~~ \notag \\
    S_{xx} =
    \iint_{\mathrm{wing}} x^2~ \mathrm{d}x \mathrm{d}y, ~~~
    S_{yy} =
    \iint_{\mathrm{wing}} y^2~ \mathrm{d}x \mathrm{d}y, ~~~
    S_{xy} =
    \iint_{\mathrm{wing}} x y~ \mathrm{d}x \mathrm{d}y \notag \\
    S_{s} =
    \iint_{\mathrm{wing}} s(x)~ \mathrm{d}x \mathrm{d}y, ~~~
    S_{sx}  =
    \iint_{\mathrm{wing}} s(x)x~ \mathrm{d}x \mathrm{d}y, ~~~
    S_{y|y|}  = 
    \iint_{\mathrm{wing}} y|y|~ \mathrm{d}x \mathrm{d}y,  ~~~ \notag \\ 
    S_{y_m s} =
    \iint_{\mathrm{wing}} \frac{1}{2}(y_{\mathrm{WT}} + y_{\mathrm{WR}}) s(x)  ~ \mathrm{d}x \mathrm{d}y \notag 
  \end{gather}

\end{widetext}
%
The functions with the subscript ``ch'' represent the force generated by the velocity in the chordwise plane $(v_{\mathrm{w},x}, 0, v_{\mathrm{w},z})$; 
$f_{\mathrm{tc,ch,bd}}$, $f_{\mathrm{tc,ch,cp}}$, and $f_{\mathrm{tc,ch,fl}}$
are the translational lifts; 
$f_{\mathrm{tc,ch,Wg}}$
is the Wagner effect of the lift; 
$f_{\mathrm{td,ch,bd}}$, $f_{\mathrm{td,ch,cp}}$, and $f_{\mathrm{td,ch,fl}}$ 
are the translational drags; 
$f_{\mathrm{td,ch,Wg}}$
is the Wagner effect of the drag; 
$f_{\mathrm{rc,ch,bd}}$ and $f_{\mathrm{rc,ch,fl}}$
are the rotational circulation lifts; and 
$f_{\mathrm{rc,ch,Wg}}$
is the rotational Wagner effect. The velocities of the lift and drag were decomposed into the body velocity term (with the subscript ``bd''), flapping velocity term (with the subscript ``fl''), and their coupling term (with the subscript ``cp'') to consider the advance ratio effect. The expressions for the rotational drag 
$f_{\mathrm{td,ch}}$
and added mass 
$f_{\mathrm{am,trans}}$
and 
$f_{\mathrm{am,rot}}$
are the same as those used in the conventional QSM.

The forces generated by the velocity in the spanwise plane (spanwise kinematics; $(0, v_{\mathrm{w},y}, v_{\mathrm{w},z})$) were formulated in the same manner (functions with subscript ``sp''). In Eq. \ref{eq:new_QSM}, 
$f_{\mathrm{tc,sp,bd}}$, $f_{\mathrm{tc,sp,cp}}$, and $f_{\mathrm{tc,sp,fl}}$
are the translational circulation lifts; 
$f_{\mathrm{tc,sp,Wg}}$
is the Wagner effect of the lift; 
$f_{\mathrm{td,sp,bd}}$, $f_{\mathrm{td,sp,cp}}$, and $f_{\mathrm{td,sp,fl}}$
are the translational drags; 
$f_{\mathrm{td,sp,Wg}}$
is the Wagner effect of the drag; 
$f_{\mathrm{rc,sp,bd}}$ and $f_{\mathrm{rc,sp,fl}}$
are the rotational circulation lifts; and 
$f_{\mathrm{rc,sp,Wg}}$
is the rotational Wagner effect. The rotational drag 
$f_{\mathrm{td,sp}}$
caused by $\omega_{\mathrm{w},x}$ was also included to improve the aerodynamic force estimation accuracy. The coefficients $\bm{\beta}$ and $\bm{\gamma}$ were optimized in the same way as in the conventional QSM.

The coefficients of the translational circulation lift, $C_{L,\mathrm{ch,bd}}$, $C_{L,\mathrm{ch,cp}}$, $C_{L,\mathrm{ch,fl}}$, $C_{L,\mathrm{sp,bd}}$, $C_{L,\mathrm{sp,cp}}$, and $C_{L,\mathrm{sp,fl}}$ are expressed as
\begin{align}
  C_{L,\mathrm{ch,bd}} &= C_L (\alpha, \beta_{\mathrm{tc,ch,bd1}}, \beta_{\mathrm{tc,ch,bd2}}) \notag \\
  &\quad \times \exp(\gamma_{\mathrm{tc,ch1}} Re_x^{\gamma_{\mathrm{tc,ch2}}}), \notag \\
  C_{L,\mathrm{ch,cp}} &= C_L (\alpha, \beta_{\mathrm{tc,ch,cp1}}, \beta_{\mathrm{tc,ch,cp2}}) \notag \\
  &\quad \times \exp(\gamma_{\mathrm{tc,ch1}} Re_x^{\gamma_{\mathrm{tc,ch2}}}), \notag \\
  C_{L,\mathrm{ch,fl}} &= C_L (\alpha, \beta_{\mathrm{tc,ch,fl1}}, \beta_{\mathrm{tc,ch,fl2}}) \notag \\
  &\quad \times \exp(\gamma_{\mathrm{tc,ch1}} Re_x^{\gamma_{\mathrm{tc,ch2}}}), \notag \\
  C_{L,\mathrm{sp,bd}} &= C_L (\alpha, \beta_{\mathrm{tc,sp,bd1}}, \beta_{\mathrm{tc,sp,bd2}}) \notag \\
  &\quad \times \exp(\gamma_{\mathrm{tc,sp1}} Re_y^{\gamma_{\mathrm{tc,sp2}}}), \notag \\
  C_{L,\mathrm{sp,cp}} &= C_L (\alpha, \beta_{\mathrm{tc,sp,cp1}}, \beta_{\mathrm{tc,sp,cp2}}) \notag \\
  &\quad \times \exp(\gamma_{\mathrm{tc,sp1}} Re_y^{\gamma_{\mathrm{tc,sp2}}}), \notag \\
  C_{L,\mathrm{sp,fl}} &= C_L (\alpha, \beta_{\mathrm{tc,sp,fl1}}, \beta_{\mathrm{tc,sp,fl2}}) \notag \\
  &\quad \times \exp(\gamma_{\mathrm{tc,sp1}} Re_y^{\gamma_{\mathrm{tc,sp2}}}). 
\end{align}
The coefficients of the translational drag 
$C_{D,\mathrm{ch,bd}}$, $C_{D,\mathrm{ch,cp}}$, $C_{D,\mathrm{ch,fl}}$, $C_{D,\mathrm{sp,bd}}$, $C_{D,\mathrm{sp,cp}}$, and $C_{D,\mathrm{sp,fl}}$
can be derived similarly. The translational drag in the data-driven QSM can be expressed as follows:
\begin{align}
  F_{\mathrm{td}} &=
  \frac{1}{2} \rho (v_{\mathrm{w},x}^2 + v_{\mathrm{w},z}^2) S_{\mathrm{w}} \beta_{\mathrm{ch}} \sin^2(\alpha_x) \sin(\alpha_x) \notag \\
  &\quad~ \times \exp (\gamma_{\mathrm{td,ch1}} Re_x^{\gamma_{\mathrm{td,ch2}}}) \notag \\
  &\quad + \frac{1}{2} \rho (v_{\mathrm{w},y}^2 + v_{\mathrm{w},z}^2) S_{\mathrm{w}} \beta_{\mathrm{sp}} \sin^2(\alpha_y) \sin(\alpha_y) \notag \\
  &\quad ~ \times \exp (\gamma_{\mathrm{td,sp1}} Re_y^{\gamma_{\mathrm{td,sp2}}}).
\end{align}
The velocity terms were combined for simplicity. When the spanwise kinematic velocity is zero ($v_{\mathrm{w},y} = 0$), $F_{\mathrm{td}}$ can be rewritten as
\begin{align}
  F_{\mathrm{td}} &= 
  \frac{1}{2}\rho (v_{\mathrm{w},x}^2 + v_{\mathrm{w},z}^2) S_{\mathrm{w}} \beta_{\mathrm{ch}} \sin^2(\alpha_x) \sin(\alpha_x) \notag \\
  &\quad ~ \times \exp (\gamma_{\mathrm{td,ch1}} Re_x^{\gamma_{\mathrm{td,ch2}}}) \notag \\
  &\quad + \frac{1}{2} \rho (v_{\mathrm{w},z}^2) S_{\mathrm{w}} \beta_{\mathrm{sp}}  \notag \\
  &\quad ~ \times \exp (\gamma_{\mathrm{td,sp1}} Re_y^{\gamma_{\mathrm{td,sp2}}}) \notag \\
  & =\frac{1}{2} \rho (v_{\mathrm{w},x}^2 + v_{\mathrm{w},z}^2) S_{\mathrm{w}} \beta_{\mathrm{ch}} \sin^2(\alpha_x) \sin(\alpha_x) \notag \\
  &\quad ~ \times \exp (\gamma_{\mathrm{td,ch1}} Re_x^{\gamma_{\mathrm{td,ch2}}}) \notag \\
  &\quad+ \frac{1}{2} \rho (v_{\mathrm{w},x}^2 + v_{\mathrm{w},z}^2) S_{\mathrm{w}} \beta_{\mathrm{sp}} \sin(\alpha_x) \sin(\alpha_x) \notag \\
  &\quad ~ \times \exp (\gamma_{\mathrm{td,sp1}} Re_y^{\gamma_{\mathrm{td,sp2}}}). 
\end{align}
Therefore, the coefficients of chordwise translational drag can be expressed as follows:
\begin{align}
  C_{D,\mathrm{ch,bd}} &= 
  \beta_{\mathrm{td,ch,bd}} \sin^2(\alpha) 
  \exp (\gamma_{\mathrm{td,ch1}} Re_x^{\gamma_{\mathrm{td,ch2}}}) \notag \\
  & \quad + \beta_{\mathrm{td,sp,bd}} \sin(\alpha) 
  \exp (\gamma_{\mathrm{td,sp1}} Re_y^{\gamma_{\mathrm{td,sp2}}}), \notag \\
  C_{D,\mathrm{ch,cp}} &= 
  \beta_{\mathrm{td,ch,cp}} \sin^2(\alpha) 
  \exp (\gamma_{\mathrm{td,ch1}} Re_x^{\gamma_{\mathrm{td,ch2}}}) \notag \\
  & \quad + \beta_{\mathrm{td,sp,cp}} \sin(\alpha) 
  \exp (\gamma_{\mathrm{td,sp1}} Re_y^{\gamma_{\mathrm{td,sp2}}}), \notag \\
  C_{D,\mathrm{ch,fl}} &= 
  \beta_{\mathrm{td,ch,fl}} \sin^2(\alpha) 
  \exp (\gamma_{\mathrm{td,ch1}} Re_x^{\gamma_{\mathrm{td,ch2}}}) \notag \\
  & \quad + \beta_{\mathrm{td,sp,fl}} \sin(\alpha) 
  \exp (\gamma_{\mathrm{td,sp1}} Re_y^{\gamma_{\mathrm{td,sp2}}}). 
\end{align}
Similarly, the coefficients of the spanwise translational drag can be expressed as follows:
\begin{align}
  C_{D,\mathrm{sp,bd}} &= 
  \beta_{\mathrm{td,sp,bd}} \sin^2(\alpha)
  \exp (\gamma_{\mathrm{td,sp1}} Re_y^{\gamma_{\mathrm{td,sp2}}}) \notag \\
  &\quad + \beta_{\mathrm{td,ch,bd}} \sin(\alpha)
  \exp (\gamma_{\mathrm{td,ch1}} Re_x^{\gamma_{\mathrm{td,ch2}}}), \notag \\
  C_{D,\mathrm{sp,cp}} &= 
  \beta_{\mathrm{td,sp,cp}} \sin^2(\alpha)
  \exp (\gamma_{\mathrm{td,sp1}} Re_y^{\gamma_{\mathrm{td,sp2}}}) \notag \\
  &\quad + \beta_{\mathrm{td,ch,cp}} \sin(\alpha)
  \exp (\gamma_{\mathrm{td,ch1}} Re_x^{\gamma_{\mathrm{td,ch2}}}), \notag \\
  C_{D,\mathrm{sp,fl}} &= 
  \beta_{\mathrm{td,sp,fl}} \sin^2(\alpha)
  \exp (\gamma_{\mathrm{td,sp1}} Re_y^{\gamma_{\mathrm{td,sp2}}}) \notag \\
  &\quad + \beta_{\mathrm{td,ch,fl}} \sin(\alpha)
  \exp (\gamma_{\mathrm{td,ch1}} Re_x^{\gamma_{\mathrm{td,ch2}}}). 
\end{align}

\begin{figure}[tbp]
    \centering
    \includegraphics[width=0.48\textwidth]{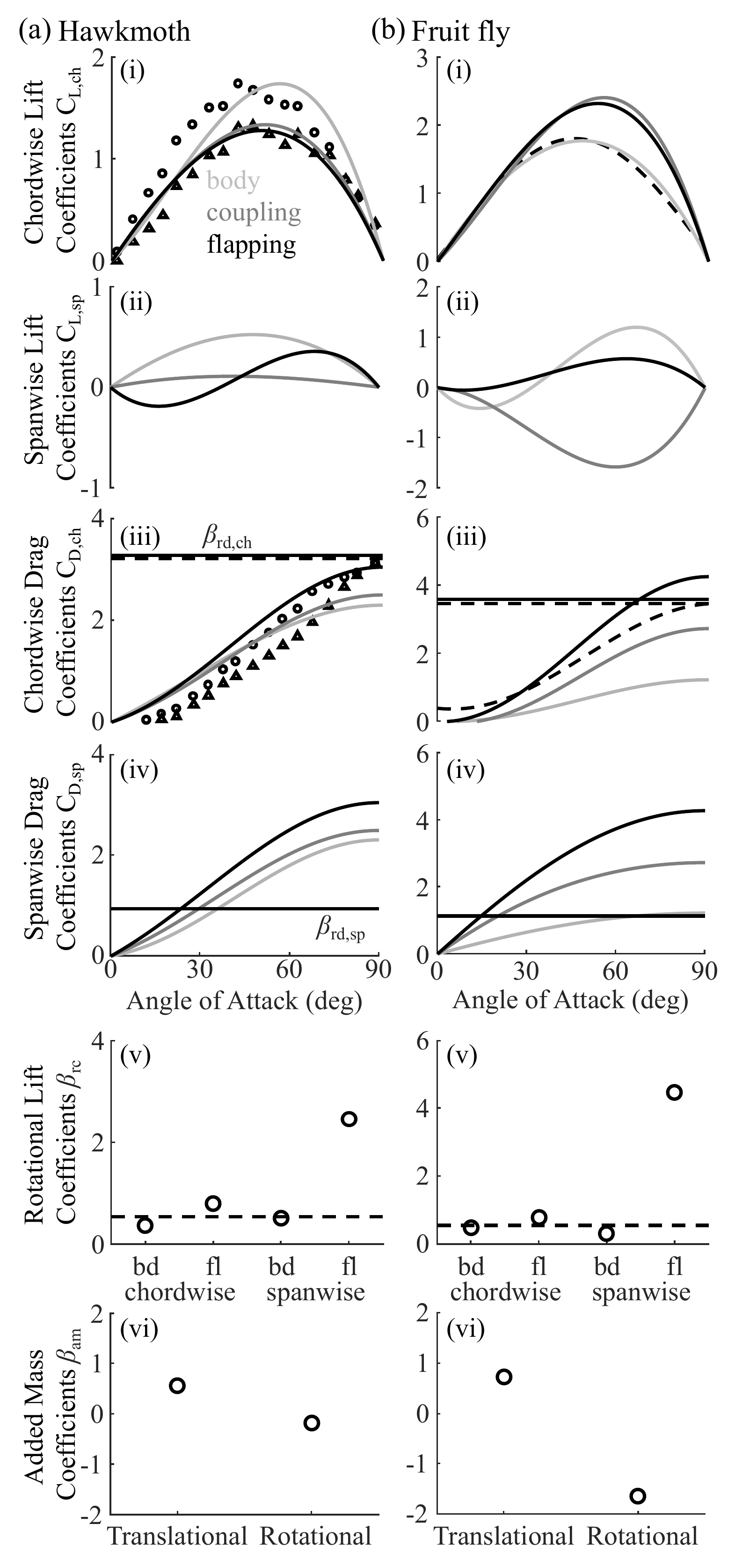}
    \caption{
    Force coefficients for the QSM of the hawkmoth (a) and fruit fly (b). (i-iv) Effect of the angle of attack on the coefficient for the (i) chordwise lift, (ii) spanwise lift, (iii) chordwise drag, and (iv) spanwise drag.
    The experimental values are shown for comparisons:
    lift and drag coeffecients measured in \cite{Usherwood2002-op} are shown in (a)-(i, iii) by the circles (``early'' condition) and triangles (``steady'' condition), and those of \cite{Dickinson1999-ug} are shown in (b)-(i, iii) by dashed curves.
    Horizontal dashed lines in (a, b)-(iii) indicate the maximum value of each drag coeffecient.
    (v) Coefficients for the rotational lift. 
    bd and fl represent body and flapping respectively. 
    Dashed lines are the rotational lift coefficients in \cite{Sane2002-ch}.
    (vi) Coefficients for the force due to translational and rotational added mass.
    }
    \label{fig:130_coefficients_DDQSM}
\end{figure}

%
\subsection{\label{subsec:App_coeffecients}Coeffecients}
The chordwise lift and drag coefficients 
$C_{L,\mathrm{ch,bd}}$, $C_{L,\mathrm{ch,cp}}$, $C_{L,\mathrm{ch,fl}}$, $C_{D,\mathrm{ch,bd}}$, $C_{D,\mathrm{ch,cp}}$, and $C_{D,\mathrm{ch,fl}}$
are in close agreement with those obtained from experiments with revolving wings at the scale of the fruit fly and hawkmoth \cite{Dickinson1999-ug,Usherwood2002-op} (Figs. \ref{fig:130_coefficients_DDQSM}(a, b)-(i, iii)). The magnitudes of the coefficients associated with the effects of the body velocity, flapping velocity, and their coupling term vary, reflecting the effect of the advance ratio. The spanwise lift coefficients 
$C_{L,\mathrm{sp,bd}}$, $C_{L,\mathrm{sp,cp}}$, and $C_{L,\mathrm{sp,fl}}$
can be negative depending on the angle of attack (Figs. \ref{fig:130_coefficients_DDQSM}(a)ii and (b)ii). This is likely because the rotation axis in the wing-normal direction ($z_{\mathrm{w}}$-axis) is positioned between the leading and trailing edges. In this configuration, the spanwise kinematic velocity ($\omega_z x$) points in opposite directions at the leading and trailing edges, occasionally resulting in a negative lift. The spanwise drag coefficients 
$C_{D,\mathrm{sp,bd}}$, $C_{D,\mathrm{sp,cp}}$, and $C_{D,\mathrm{sp,fl}}$
exhibit trends similar to those of the chordwise drag coefficients 
$C_{D,\mathrm{ch,bd}}$, $C_{D,\mathrm{ch,cp}}$, and $C_{D,\mathrm{ch,fl}}$
(Figs. \ref{fig:130_coefficients_DDQSM}(a)iv and (b)iv).

At the scale of the fruit fly, the coefficients of rotational circulation (Figs. \ref{fig:130_coefficients_DDQSM}(a)v and (b)v) align with those reported in a previous study \cite{Sane2002-ch}, except for $\beta_{\mathrm{rc,sp,fl}}$, which shows markedly different values. This is likely because the rotation axis in the wing-normal direction is located near the middle of the chord. The coefficients of rotational drag (horizontal solid lines in Figs. \ref{fig:130_coefficients_DDQSM}(a, b)-(iii, iv)) are consistent with those reported in previous studies (horizontal broken lines, \cite{Dickinson1999-ug,Usherwood2002-op}). For the force due to added mass, the coefficients for translational acceleration ($a_{\mathrm{w},z}$), $\beta_{\mathrm{am,trans}}$, exhibit a value close to 1, as expected. The coefficient for rotational acceleration ($\psi_{\mathrm{w},y}$), $\beta_{\mathrm{am,rot}}$, varies with the location of the rotation axis; specifically, $\beta_{\mathrm{am,rot}}$ increases in the positive direction as the rotation axis moves toward the leading edge. In this study, the wing hinge of the fruit fly was positioned more posteriorly compared to that of the hawkmoth (Figs. \ref{fig:20_wingkin_morphology}(d) and (e)), resulting in larger negative values of $\beta_{\mathrm{am,rot}}$ for the fruit fly wing.

%
\subsection{\label{subsec:App_formulationWagner}Formulation of the Wagner Effect}
van Veen et al. \cite{van-Veen2023-dq} proposed a new interpretation of the Wagner effect for flapping wings as an interaction between the velocity and acceleration. They modeled the delay of development and decay of translational circulation $\Gamma_{\mathrm{tc}}$ as being proportional to 
$\sqrt{c a}/v$
($c$ is the chord length, $a$ is the acceleration, and $v$ is the velocity), the ratio between the timescales of velocity and acceleration, and numerically demonstrated its validity. Considering this effect, the translational circulation lift $\mathrm{d} F_{\mathrm{tc}}$ can be expressed as
\begin{align}
  \Gamma_{\mathrm{tc}}(t) &= \Gamma_{\mathrm{qs}}(t) \left( 1 - k \frac{\sqrt{ca(t)}}{v(t)} \right), \label{eq:circ_with_wagner_orig} \\
  \mathrm{d} F_{\mathrm{tc}} &= \rho v(t) \Gamma_{\mathrm{tc}}(t) \mathrm{d} y, \label{eq:tc_with_wagner_orig}
\end{align}
where $\Gamma_{\mathrm{qs}}(t)$ is a circulation based on the quasi-steady assumption, and $k$ is a constant value. 
%
%
%
We assumed that the model in \cite{van-Veen2023-dq} based on constant acceleration holds within small timescales, thereby extending it to accommodate the varying accelerations of flapping wings.
In other words, the delay of the development and decay of circulation in small timestep $\Delta t$ is proportional to $\sqrt{ca}/v$, as shown below.
\begin{align}
  &\Gamma_{\mathrm{tc}}(t+\Delta t) - \Gamma_{\mathrm{tc}}(t) \notag \\
  & \quad =
  \left( \Gamma_{\mathrm{qs}}(t+\Delta t) - \Gamma_{\mathrm{qs}}(t) \right) 
  \left( 1 - k \frac{\sqrt{ca(t)}}{v(t)} \right). 
\end{align}
Taking the limit $\Delta t \to 0$, the circulation $\Gamma_{\mathrm{tc}}(t)$ at time $t$ can be given by
\begin{equation}
  \Gamma_{\mathrm{tc}}(t)=\Gamma_{\mathrm{tc}}(t_0)
  + \int_{t_0}^{t} \frac{\mathrm{d}\Gamma_{\mathrm{qs}}(t)}{\mathrm{d} \tau}
  \left( 1 - k \frac{\sqrt{ca(t)}}{v(t)} \right) \mathrm{d}\tau,
\end{equation}
where $t_0$ is the time at which acceleration or deceleration starts $(a(t_0) = 0)$. We can further rewrite this equation as:
\begin{align}
  \Gamma_{\mathrm{tc}}(t) &= \Gamma_{\mathrm{tc}}(t_0)
  + \Gamma_{\mathrm{qs}}(t) \left( 1 - k \frac{\sqrt{ca(t)}}{v(t)} \right) \notag \\
  &\quad - \Gamma_{\mathrm{qs}}(t_0) \left( 1 - k \frac{\sqrt{ca(t_0)}}{v(t_0)} \right) \notag \\
  &\quad - \int_{t_0}^{t} \Gamma_{\mathrm{qs}}(\tau) 
  \frac{\mathrm{d}}{\mathrm{d} \tau} \left( 1 - k \frac{\sqrt{ca(t)}}{v(t)} \right) \mathrm{d}\tau. 
\end{align}
In addition, there exists a $t_{\mathrm{eq}}$ such that $t_0 \leq t_{\mathrm{eq}} \leq t$, for which the following holds:
{\small
  \begin{align}
    \Gamma_{\mathrm{tc}}(t) &= \Gamma_{\mathrm{tc}}(t_0)
    + \Gamma_{\mathrm{qs}}(t) \left( 1 - k \frac{\sqrt{ca(t)}}{v(t)} \right) \notag \\
    &\quad - \Gamma_{\mathrm{qs}}(t_0) \left( 1 - k \frac{\sqrt{ca(t_0)}}{v(t_0)} \right) \notag \\
    &\quad - \Gamma_{\mathrm{qs}}(t_{\mathrm{eq}})
    \left( 
      \left( 1 - k \frac{\sqrt{ca(t)}}{v(t)} \right)
      -\left( 1 - k \frac{\sqrt{ca(t_0)}}{v(t_0)} \right)
    \right). \notag \\
    & 
  \end{align}
}
%
Here, $\Gamma_{\mathrm{qs}}(t_0)= \Gamma(t_0)$, $a(t_0) = 0$; therefore,
\begin{align}
  \Gamma_{\mathrm{tc}}(t) &= 
  \Gamma_{\mathrm{qs}}(t) \left( 1 - k \frac{\sqrt{ca(t)}}{v(t)} \right) \notag \\
  &+ \Gamma_{\mathrm{qs}}(t_{\mathrm{eq}}) k \frac{\sqrt{ca(t)}}{v(t)} \notag \\
  &= \Gamma_{\mathrm{qs}}(t) 
  + k \left( \Gamma_{\mathrm{qs}}(t_{\mathrm{eq}}) - \Gamma_{\mathrm{qs}}(t) \right) \frac{\sqrt{ca(t)}}{v(t)}. \notag \\
  & 
\end{align}
When considering deceleration, we only need to replace $\sqrt{a}$ with $\mathrm{sign}(a) \sqrt{|a|}$ as in \cite{van-Veen2023-dq}. The above formula indicates that the circulation at time $t$ can be expressed by a linear interpolation of the quasi-steady circulation at times $t$ and $t_{\mathrm{eq}}$. The quasi-steady translational circulation can be expressed as:
\begin{equation}
  \Gamma_{\mathrm{qs}}(t) = \frac{1}{2}cv(t)C_{\mathrm{F}}(t), 
\end{equation}
therefore,
\begin{align}
  \Gamma_{\mathrm{tc}}(t) &= \frac{1}{2}cv(t)C_{\mathrm{F}}(t) \notag \\
  &+ \frac{1}{2}c k
  \left(
    v(t_{\mathrm{eq}}) C_{\mathrm{F}}(t_{\mathrm{eq}})
    - v(t) C_{\mathrm{F}}(t)
  \right)
  \frac{\sqrt{ca(t)}}{v(t)}, \notag \\
  &
\end{align}
thus, the translational circulation force can be written as
\begin{align}
  \mathrm{d} F_{\mathrm{tc}} &= \rho v(t) \Gamma_{\mathrm{tc}}(t) \mathrm{d}y \notag \\
  &= \frac{1}{2} \rho v(t)^2 C_{\mathrm{F}}(t) c \mathrm{d} y \notag \\
  &\quad + \frac{1}{2} \rho k \left(
    v(t_{\mathrm{eq}}) C_{\mathrm{F}}(t_{\mathrm{eq}})
    - v(t) C_{\mathrm{F}}(t)
  \right)
  \sqrt{c a(t)} ~ c \mathrm{d} y, \notag \\
  & \label{eq:tc_with_wagner_revised}
\end{align}
where $C_{\mathrm{F}}$ is the force coefficient and $k$ is a constant. The first and second terms correspond to the quasi-steady and Wagner effects, respectively. If we assume that the acceleration and angle of attack are constant, then Eq. \ref{eq:tc_with_wagner_revised} becomes equivalent to the formulation proposed by van Veen et al. \cite{van-Veen2023-dq} (Eqs. \ref{eq:circ_with_wagner_orig} and \ref{eq:tc_with_wagner_orig}).

Assuming the formulation above also holds for the rotational circulation $\Gamma_{\mathrm{rc}}(t)$, the rotational circulation can be given by
\begin{align}
  \Gamma_{\mathrm{rc}}(t) &= 
  \beta_{\mathrm{rc}} \omega(t) c^2 \notag \\
  &\quad + k' \left(
    \omega(t_{\mathrm{eq}})-\omega(t)
  \right) \frac{ \sqrt{c a(t)} }{v(t)} c^2
\end{align}
and the rotational circulation lift $F_{\mathrm{tc}}(t)$ is
\begin{align}
  \mathrm{d} F_{\mathrm{rc}} &= \rho v(t) \Gamma_{\mathrm{rc}}(t) \mathrm{d}y \notag \\
  &= \beta_{\mathrm{rc}} \rho v(t) \omega(t) c^2 \mathrm{d}y \notag \\
  &\quad + k' \rho \left(
    \omega(t_{\mathrm{eq}})-\omega(t)
  \right) \sqrt{c a(t)} c^2 \mathrm{d}y
\end{align}
where the second term represents the Wagner effect (we call it the ``rotational Wagner effect'' here). In our study, based on \cite{Bayiz2021-yx}, we set $t_{\mathrm{eq}}$ to one-eighth the stroke cycle before time $t$.



\bibliography{bibliography_aligned.bib, bibliography_others.bib}

\clearpage
%

\end{document}